\documentclass[12pt]{article}

\usepackage{newtxtext,newtxmath}

\usepackage{graphicx}

\usepackage[letterpaper,margin=1in]{geometry}

\renewenvironment{abstract}
	{\quotation}
	{\endquotation}

\date{}

\makeatletter
\renewcommand{\fnum@figure}{\textbf{Figure \thefigure}}
\renewcommand{\fnum@table}{\textbf{Table \thetable}}
\makeatother

\usepackage{scicite}

\usepackage{url}

\def\aj{Astron. J.}                   
\def\araa{Annu. Rev. Astron. Astrophys.}             
\def\apj{Astrophys. J.}                 
\def\apjl{Astrophys J. Lett.}                
\def\apjs{Astrophys. J. Suppl. Ser.}               
\def\aap{Astron. Astrophys.}                
\def\mnras{Mon. Not. R. Astron. Soc.}             
\def\nat{Nature}              
\def\natast{Nature Astron.}
\def\pasp{Publ. Astron. Soc. Pac.}               

\newcommand{\kms}{km~s$^{-1}$}
\newcommand{\msol}{~${\rm M}_{\odot}$}
\usepackage[T1]{fontenc}

\def\scititle{
	A stellar dynamical mass measurement of an inactive black hole at redshift 2
}
\title{\bfseries \boldmath \scititle}

\author{
    \normalsize
    Andrew~B.~Newman$^{1,2\ast}$,
    Meng~Gu$^{3\dagger\ddag}$,
    Sirio~Belli$^4$,
    Richard~S.~Ellis$^5$,
    Sai Gangula$^{2,1}$,\and
    \normalsize Jenny~E.~Greene$^6$,
    Jonelle~L.~Walsh$^{7,8}$,
    Sherry~H.~Suyu$^{9,10}$,
    Sebastian~Ertl$^{10,9}$,
    Gabriel~Caminha$^{10,9}$,\and
    \normalsize Giovanni~Granata$^{11,12,13}$,
    Claudio~Grillo$^{11,14}$,
    Stefan~Schuldt$^{11,14\S\parallel}$,
    Tania~M.~Barone$^{15}$,
    Simeon~Bird$^{16}$,\and
    \normalsize  Karl~Glazebrook$^{15}$,
    Marziye~Jafariyazani$^{17,18}$,
    Mariska~Kriek$^{19}$,
    Allison~Matthews$^1$,\and
    \normalsize Takahiro~Morishita$^{20\P}$,
    Themiya~Nanayakkara$^{15}$,
    Justin~D.~R.~Pierel$^{21}$,
    Ana~Acebr\'on$^{22,14}$,\and
    \normalsize Pietro~Bergamini$^{11,23}$,
    Sangjun~Cha$^{24}$,
    Jose~M.~Diego$^{22}$,
    Nicholas~Foo$^{25}$,
    Brenda~Frye$^{26}$,\and
    \normalsize Yoshinobu~Fudamoto$^{27,26}$,
    M.~James~Jee$^{24,28}$,
    Patrick~S.~Kamieneski$^{25\dagger\dagger}$,
    Anton~M.~Koekemoer$^{21}$,\and
    \normalsize Asish~K.~Meena$^{29\ddag\ddag}$,
    Shun~Nishida$^{30}$,
    Masamune~Oguri$^{27,30}$,
    Piero~Rosati$^{12,23}$,
    Adi~Zitrin$^{29}$\and
    \small$^1$Observatories, Carnegie Institution for Science, Pasadena, CA, USA.\and
    \small$^2$Department of Physics and Astronomy, University of Southern California, Los Angeles, CA, USA.\and
    \small$^3$Department of Physics, The University of Hong Kong, Hong Kong, China.\and
    \small$^4$Dipartimento di Fisica e Astronomia, Università di Bologna, Bologna, Italy.\and
    \small$^5$Department of Physics \& Astronomy, University College London, London, UK.\and
    \small$^6$Department of Astrophysical Sciences, Princeton University, Princeton, NJ, USA.\and
    \small$^7$George P. and Cynthia Woods Mitchell Institute for Fundamental Physics and Astronomy,\and
    \small Texas A\&M University, College Station, TX, USA.\and
    \small$^8$Department of Physics and Astronomy, Texas A\&M University, College Station, TX, USA.\and
    \small$^{9}$School of Natural Sciences, Technical University of Munich, Garching, Germany.\and
    \small$^{10}$Max-Planck-Institut f\"ur Astrophysik, Garching, Germany.\and
    \small$^{11}$Dipartimento di Fisica, Università degli Studi di Milano, Milano, Italy.\and
    \small$^{12}$Dipartimento di Fisica e Scienze della Terra, Università degli Studi di Ferrara, Ferrara, Italy.\and
    \small$^{13}$Institute of Cosmology and Gravitation, University of Portsmouth, Portsmouth, UK.\and
    \small$^{14}$Instituto di Astrofisica Spaziale e Fisica cosmica Milano, Istituto Nazionale di Astrofisica, Milano, Italy.\and
    \small$^{15}$Centre for Astrophysics and Supercomputing, Swinburne University of Technology, \and
    \small Hawthorn, Victoria, Australia.\and
    \small$^{16}$Department of Physics \& Astronomy, University of California Riverside, Riverside, CA, USA.\and
    \small$^{17}$SETI Institute, Mountain View, CA, USA.\and
    \small$^{18}$NASA Ames Research Center, Moffett Field, CA, USA.\and
    \small$^{19}$Leiden Observatory, Leiden University, Leiden, The Netherlands.\and
    \small$^{20}$Infrared Processing and Analysis Center, California Institute of Technology, Pasadena, CA, USA.\and
    \small$^{21}$Space Telescope Science Institute, Baltimore, MD, USA.\and
    \small$^{22}$Instituto de F\'isica de Cantabria, Santander, Spain.\and
    \small$^{23}$Osservatorio di Astrofisica e Scienza dello Spazio di Bologna,\and
    \small Istituto Nazionale di Astrofisica, Bologna, Italy.\and
    \small$^{24}$Department of Astronomy, Yonsei University, Seoul, Korea.\and
    \small$^{25}$School of Earth and Space Exploration, Arizona State University, Tempe, AZ, USA.\and
    \small$^{26}$Department of Astronomy and Steward Observatory, University of Arizona, Tucson, AZ, USA.\and
    \small$^{27}$Center for Frontier Science, Chiba University, Chiba, Japan.\and
    \small$^{28}$Department of Physics and Astronomy, University of California Davis, Davis, CA, USA.\and
    \small$^{29}$Department of Physics, Ben-Gurion University of the Negev, Be'er-Sheva, Israel.\and
    \small$^{30}$Department of Physics, Graduate School of Science, Chiba University, Chiba, Japan.\and
    \small$^\ast$Corresponding author. Email: anewman@carnegiescience.edu\and
    \small$^\dagger$Present address: Department of Astronomy, Tsinghua University, Beijing, China.\and
    \small$^\ddag$Present address: Hong Kong Institute for Astronomy \& Astrophysics,\and\small The University of Hong Kong, Hong Kong, China.\and
	\small$^\S$Present address: Finnish Centre for Astronomy with ESO, Turku, Finland.\and
	\small$^\parallel$Present address: Department of Physics, University of Helsinki, Helsinki, Finland.\and
	\small$^\P$Present address: Astronomical Institute, Tohoku University, Sendai, Japan.\and
	\small${}^{\dagger\dagger}$Present address: Department of Space, Earth \& Environment,\and\small Chalmers University of Technology, Gothenburg, Sweden.\and
	\small${}^{\ddag\ddag}$Present address: Department of Physics, Indian Institute of Science, Bengaluru, India.
}

\begin{document} 

\maketitle

\begin{abstract} \bfseries \boldmath
Supermassive black holes and their host galaxies grow together over time, producing correlations between the black hole mass and various galaxy properties. Determining the evolution of these correlations requires precise measurements of the masses of distant black holes. We observe the gravitationally lensed quiescent galaxy MRG-M0138, at redshift 1.95, using JWST integral field spectroscopy to spatially resolve the kinematics of stars within the black hole's sphere of influence. By using a foreground lens model and fitting stellar dynamical models, we determine the mass of its inactive black hole, $M_{\bullet}=6.0^{+2.1}_{-1.7}\times10^9$ solar masses. Comparing this measurement to local galaxies, we find that $M_{\bullet}$ is higher than expected given the galaxy's bulge mass, but consistent with the correlation with stellar velocity dispersion.
\end{abstract}

\noindent
The growths of supermassive black holes and their host galaxies are intertwined, and observations have shown that the black hole mass $M_{\bullet}$ correlates with various host galaxy properties, such as the bulge mass and the stellar velocity dispersion \cite{KH13,McConnell13,Greene20}. It is unclear whether these correlations evolved over the history of the Universe. The masses of distant black holes have been estimated in active galactic nuclei (AGNs), where accretion onto the black hole powers substantial energy output. The accreting material irradiates fast-moving gas clouds in the surrounding broad-line region, enabling their kinematics to be measured when the viewing angle is favorable. The angular size of the broad-line region at redshifts $z \gtrsim 2$ is $\sim$100 microarcseconds or less, which has been spatially resolved in only two cases \cite{Abuter24,Gravity2026}. Most measurements instead apply mass estimators to emission line widths and luminosities measured in unresolved spectra \cite{Greene05}. However, this approach relies on estimates of the size and structure of the broad-line region, resulting in large uncertainties of up to an order of magnitude \cite{Abuter24,Berternes25,Gravity2026}.
 
Measuring the properties of the host galaxies of AGNs is also challenging, especially for the most massive known black holes in the distant Universe. Those black holes are found by locating quasars, luminous AGNs that greatly outshine the stars in their hosts. In contrast, observing galaxies with inactive black holes permits both an investigation of the host galaxy properties and a robust measurement of the black hole mass using stellar dynamics. At $z \gtrsim 0.3$, these observations require an angular resolution of tens of milliarcseconds or better to resolve the black hole's sphere of influence, where the motions of stars are strongly influenced by the gravitational potential of the black hole. Consequently, precise black hole mass measurements are limited to distances $< 200$~megaparsecs (Mpc) \cite{Mehrgan19}. This limitation could potentially be overcome if the sphere of influence is highly magnified by gravitational lensing.

\subsubsection*{Observations and host galaxy properties}

We used JWST to observe one of five images of the quiescent galaxy MRG-M0138 (right ascension $01^{\rm h}38^{\rm m}03.17^{\rm s}$, declination $-21^{\circ}55'47.6''$) at $z=1.95$ \cite{Newman18a}, which is highly magnified by gravitational lensing caused by the foreground galaxy cluster MACS~J0138.0--2155 (Fig.~\ref{fig1}A). The galaxy is intrinsically luminous and dominated by an old stellar population with minimal dust extinction, which enables precise stellar kinematic measurements. The observations used the integral field unit (IFU) of the Near Infrared Spectrograph (NIRSpec) on JWST \cite{Jakobsen22} with two gratings, which together cover the rest-frame wavelengths $\lambda_{\rm rest} = 0.33$ to 1.07~µm at a resolving power $R \sim 1000$. We also analyzed images obtained using the Advanced Camera for Surveys on the Hubble Space Telescope (HST) and the Near Infrared Camera (NIRCam) on JWST, which span $\lambda_{\rm rest}=0.19$ to 1.5~µm in seven filters.

We interpreted the observations using a model of the gravitational lens \cite{Ertl25} and used the results of seven independently constructed lens models to estimate uncertainties in the magnification $\mu = 29^{+13}_{-11}$\cite{Suyu25} and its anisotropy \cite{methods}. Given this magnification, the point spread function (PSF) of the IFU observations is equivalent to a  spatial resolution of $\sigma_{\rm psf} = 91^{+41}_{-35}$~parsecs (pc) \cite{methods}.

We reconstructed the NIRCam images in the source plane (Fig.~\ref{fig1}B), which show that the galaxy has a multi-component structure: an inclined disk contributes 62\% of the light at $\lambda_{\rm rest} \approx 0.68$~µm, with the remainder from a compact stellar bulge. We extracted spectra from the IFU data cube in 219 spatial bins [not all independent, \cite{methods}] along with matched photometry from the seven images. We used stellar population synthesis (SPS) techniques to model the spectra and photometry, producing maps of the stellar velocity $V$ (Fig.~\ref{fig1}C), stellar velocity dispersion $\sigma$ (Fig.~\ref{fig1}D), and the stellar mass and light distributions (Fig.~\ref{fig:structure}).

We find that MRG-M0138 is a massive and dense galaxy, with total stellar mass $M_{*, \rm MW} = 2.2^{+1.4}_{-0.7} \times 10^{11}$~solar masses (${\rm M}_\odot$) assuming a Milky Way (MW) stellar initial mass function (IMF) \cite{Kroupa01}. It has an effective radius $R_e = 2.73^{+0.77}_{-0.46}$~kiloparsecs (kpc), defined as the half-light semi-major axis, which is typical of quiescent galaxies of its mass and redshift \cite{vanderWel14}. It also has a high effective velocity dispersion $\sigma_e = 398 \pm 12$~\kms, which we define as the second velocity moment $V_{\rm rms} = \sqrt{V^2 + \sigma^2}$ averaged within the bulge effective radius, $R_{e, \rm bulge} = 0.83^{+0.23}_{-0.14}$~kpc. [Uncertainties are based on estimates of systematics \cite{methods}, and rms denotes root mean square.]

\subsubsection*{Black hole mass and Eddington ratio}

We modeled the stellar dynamics of the host galaxy using Jeans anisotropic modeling [JAM, \cite{Cappellari08}], a technique that has been shown to produce accurate $M_{\bullet}$ estimates in fast-rotating galaxies \cite{Thater19} like MRG-M0138. We constructed flexible dynamical models that allow us to estimate $M_{\bullet}$ after marginalizing over a range of stellar mass distributions, velocity anisotropies, and inclinations \cite{methods}. Our initial estimates of the stellar mass distributions of the bulge and disk relied on the SPS model fitting, which allowed for gradients in stellar population age, metal abundance, and dust but assumed a Milky Way IMF. The dynamical models permit this initial stellar mass distribution to vary, thereby allowing for the possibility of a non-Milky Way IMF with spatial gradients. The small gas mass in MRG-M0138 \cite{Whitaker21} is dynamically negligible \cite{methods}. The dynamical models also include dark matter, a central black hole, and gradients in the velocity anisotropy.

The dynamical models were fit to the observed $V_{\rm rms}$ map (Fig.~\ref{fig2}). We considered six model variations to test the influence of both the uncertain IMF and its possible spatial gradients, and the assumed orientation of the stellar velocity ellipsoid, which describes the anisotropy of the velocity dispersion \cite{methods}. All six model variations produce consistent estimates of $M_{\bullet}$ (Fig.~\ref{fig:mbh_posterior}). We combined these estimates using Bayesian model averaging, finding $\log(M_{\bullet} / {\rm M}_{\odot}) = 9.78^{+0.08}_{-0.12}(1\sigma~{\rm stat.}){}^{+0.11}_{-0.09}{\rm (sys.)}$; combining the statistical and systematic uncertainties in quadrature gives $\log(M_{\bullet} / {\rm M}_{\odot}) = 9.78^{+0.13}_{-0.15}$. Among the six dynamical models, we selected a fiducial model as the basis for estimating systematic uncertainties and conducting the tests described below. The largest contribution to the systematic uncertainty is the lensing magnification \cite{methods}. The radius of the black hole sphere of influence, $r_{\rm inf} = 164$~pc, is marginally resolved by a factor $r_{\rm inf} / \sigma_{\rm psf} = 1.8$, which is similar to that achieved in observations of nearby black holes with lower masses \cite{Thater19}. The fitted dynamical mass of the stellar bulge is $M_{\rm bulge} = 1.03^{+0.33}_{-0.22} \times 10^{11}$\msol, and the total dynamical stellar mass is $M_{\rm stars} = 2.54^{+0.87}_{-0.55} \times 10^{11}$\msol~(uncertainties combine $1\sigma$ statistical and systematic components).

We also considered several dynamical models that omit a black hole. These models do not match the central $V_{\rm rms}$ peak in the observations (Fig.~\ref{fig2}C-E and Fig.~\ref{fig:compare_bh_nobh}), even if an IMF gradient is allowed to increase the central mass density. Compared to the fiducial dynamical model that includes a black hole, the models with no black hole are disfavored by Bayes factors ranging from 735 to $8 \times 10^{12}$. As a further test, we replaced the black hole in the fiducial dynamical model with an extended dark mass with a free mass, size, and shape. We found that the kinematics require that the dark mass is spatially unresolved, setting an upper limit (95\%) on its radius of $28$~pc (Fig.~\ref{fig:dark_mass}), which is close to the source-plane spatial resolution of the IFU data in the most-magnified direction. The required surface density within this radius is $10^{6.8 \pm 0.4}$~\msol~pc$^{-2}$, which is implausibly high for a stellar system mimicking a black hole \cite{methods}.

We isolated the nuclear spectrum within $r \approx 100$~pc of the galaxy center to place limits on any AGN activity. Faint emission lines of singly ionized nitrogen and oxygen ([N~\textsc{ii}] and [O~\textsc{ii}]) were detected (Fig.~\ref{fig:emlinespec}), and upper limits were placed on emission from hydrogen (H$\alpha$) and doubly ionized oxygen ([O~\textsc{iii}]). The flux ratios, log([N~\textsc{ii}]~/~H$\alpha$) $> 0.75$ and log([O~\textsc{iii}]~/~[O~\textsc{ii}]) $< -0.10$ (2$\sigma$ limits), are consistent with the presence of a low-ionization nuclear emission region (LINER). The upper limit on H$\alpha$ emission constrains the AGN bolometric luminosity $L_{\rm bol} < 10^{42.8}$~erg~s${}^{-1}$  and the Eddington ratio $\lambda_{\rm Edd} < 10^{-5.1}$ \cite{methods}, which expresses the ratio of $L_{\rm bol}$ to a theoretical maximum Eddington luminosity (upper limits include the statistical $2\sigma$ range and systematic uncertainties). No X-ray emission from this galaxy is detected in an archival observation by the Chandra X-ray Observatory, which provides a consistent (but 0.6~dex weaker) limit on $L_{\rm bol}$ \cite{methods}. MRG-M0138 therefore has a similarly low Eddington ratio as is seen in typical local galaxies not classified as AGN, where the most common mode of accretion is expected to be inefficient at producing radiation \cite{Ho09}. Fig.~\ref{fig3}A compares our limit on $L_{\rm bol}$ in MRG-M0138 to galaxies in the local Universe and previous $M_{\bullet}$ measurements at $z > 1$ based on broad emission lines; the latter all have $\lambda_{\rm Edd} \gtrsim 0.01$.

\subsubsection*{Black hole scaling relations}

Our results indicate that the black hole in MRG-M0138 is as massive as those in quasars (Fig.~\ref{fig3}A). Quasars are identified based on their high bolometric luminosities; this preferentially selects the most massive black holes and makes it difficult to infer the evolution of the galaxy--black hole correlations \cite{Lauer07}. In contrast, stellar dynamical methods allow measurements of $M_{\bullet}$ in inactive galaxies like MRG-M0138, circumventing this type of selection bias. We find that the black hole mass fraction $M_{\bullet} / M_{\rm stars} = 0.024^{+0.008}_{-0.007}$ is an order of magnitude higher than the average of faint AGN at the same redshift \cite{Suh20}. It is also near the top of the range of $M_{\bullet} / M_{\rm stars}$ ratios typically observed in local early-type (elliptical and lenticular) or late-type (spiral) galaxies (Fig.~\ref{fig3}B).

Most cosmological simulations do not produce $z=2$ galaxies with both the $M_{\bullet}$ and $M_{\rm stars}$ that we measured for MRG-M0138 \cite{Habouzit21}. For example, the Astrid simulation contains 635 galaxies at $z = 2$ that match the stellar mass of MRG-M0138 (within $1\sigma$), but only two contain a black hole as massive (within $1\sigma$ or above) \cite{Ni25}. The semi-empirical model TRINITY predicts the median $M_{\bullet}$--$M_{\rm stars}$ relation at $z=2$ \cite{Zhang23}, and the black hole in MRG-M0138 is a factor of five above it, a difference of 2.6 standard deviations (our measurement uncertainty added in quadrature to the intrinsic scatter, according to TRINITY). Our results are consistent with previous indications that early black hole growth was very efficient in at least some galaxies \cite{Goulding23,Juodzbalis24,Yue24}.

The origin of the high $M_{\bullet}$ in MRG-M0138 could be related to the galaxy density, as indicated by the very high value of $\sigma_e$. MRG-M0138 is consistent with the local $M_{\bullet}$--$\sigma_e$ relation (Fig.~\ref{fig3}D). Previous studies of high-redshift galaxies that used gas emission lines as proxies for $\sigma_e$ \cite{Maiolino24} are also consistent with the local relation, implying that it does not strongly evolve. This result informs uncertain models of black hole growth; cosmological simulations predict a wide range in the amount of evolution in the $M_{\bullet}$--$\sigma_e$ relation \cite{Sijacki15,Thomas19}. Our measurement also indirectly supports some previous broad-line $M_{\bullet}$ estimates at $z > 2$: if the $M_{\bullet}$--$\sigma_e$ relation has negligible evolution, $\sigma_e$ can be used to estimate $M_{\bullet}$ at those redshifts. The resulting estimate is consistent with broad-line estimates \cite{Carnall23,Ito25} in two quiescent galaxies at $z\approx2$ and 5 (Fig.~\ref{fig3}D).

In contrast, MRG-M0138 lies above the $M_{\bullet}$--$M_{\rm bulge}$ relation of local early-type galaxies \cite{KH13} by a factor of 12 (Fig.~\ref{fig3}C), a difference of 3.3 standard deviations (our measurement uncertainty added in quadrature to the intrinsic scatter). If MRG-M0138 is typical of galaxies at $z \sim 2$, it indicates that the $M_{\bullet}$--$M_{\rm bulge}$ relation has evolved much more than the $M_{\bullet}$--$\sigma_e$ relation. The reason is probably that galaxies underwent significant morphological change, including bulge growth, after the main period of star formation and black hole growth \cite{Newman18b}. Similar conclusions on the evolution of the $M_{\bullet}$--$M_{\rm bulge}$ relation have been reached by studies of quasar hosts using statistical estimates of $M_{\rm bulge}$ \cite{Ding20}. 

\subsubsection*{Possible evolutionary pathways}

The dominant growth mechanism for the black hole in MRG-M0138 was probably accretion during optically luminous quasar phases \cite{Yu02}. If it grew at $\lambda_{\rm Edd} \sim 0.3$, which is typical of quasars \cite{Kollmeier06}, its AGN luminosity would have been $L_{\rm bol} \approx 10^{47}$~erg~s${}^{-1}$. Quasars with such luminosities often drive winds that eject gas at outflow rates of $\sim10^3$\msol~yr${}^{-1}$ \cite{Fiore17}. Our $M_{\bullet}$ measurement  therefore indicates that MRG-M0138 was likely a site of powerful AGN feedback. We cannot demonstrate a causal link between that feedback and the quenching of star formation in MRG-M0138 or its low gas fraction. However, this scenario is consistent with theoretical models that produce massive elliptical galaxies from a merger-induced starburst and quasar phase \cite{Hopkins08}. A major merger would have destroyed the stellar disks of the progenitors, but that is not incompatible with the thin stellar disk in MRG-M0138 \cite{Newman18b}. Simulations of major mergers predict that if the gas fraction is as high as observed at $z \gtrsim 2$ \cite{Tacconi18}, a substantial fraction of the gas is not consumed by the post-merger starburst and instead re-forms a gas disk, where star formation may continue \cite{Robertson06,Hopkins09}.

MRG-M0138 will likely become an elliptical galaxy by $z \sim 0$: its mass is  $M_{\rm stars} \approx 10^{11.4}$\msol, which we expect to double between $z = 2$ and 0, primarily due to mergers \cite{Brammer11}. At $z \sim 0$, this mass regime is dominated by slowly rotating, elliptical galaxies \cite{Emsellem11}. If these later mergers are gas-poor, they would cause little black hole growth \cite{Kulier15} and only modest changes to $\sigma_e$ \cite{Hilz12}. MRG-M0138 would remain consistent with the local $M_{\bullet}$--$\sigma_e$ relation (Fig.~\ref{fig3}D) but evolve towards the mean $M_{\bullet}$--$M_{\rm stars}$ relation (Fig.~\ref{fig3}B) of early-type galaxies. To form an elliptical galaxy, the mergers must redistribute stars from the stellar disk into a bulge; along with the expected mass doubling, this evolution would bring MRG-M0138 onto the local $M_{\bullet}$--$M_{\rm bulge}$ relation (Fig.~\ref{fig3}C), producing black hole and stellar masses similar to the nearby elliptical galaxy Messier 87 \cite{Oldham16,EHT19}.

\begin{figure}
    \centering
    \includegraphics[width=\linewidth]{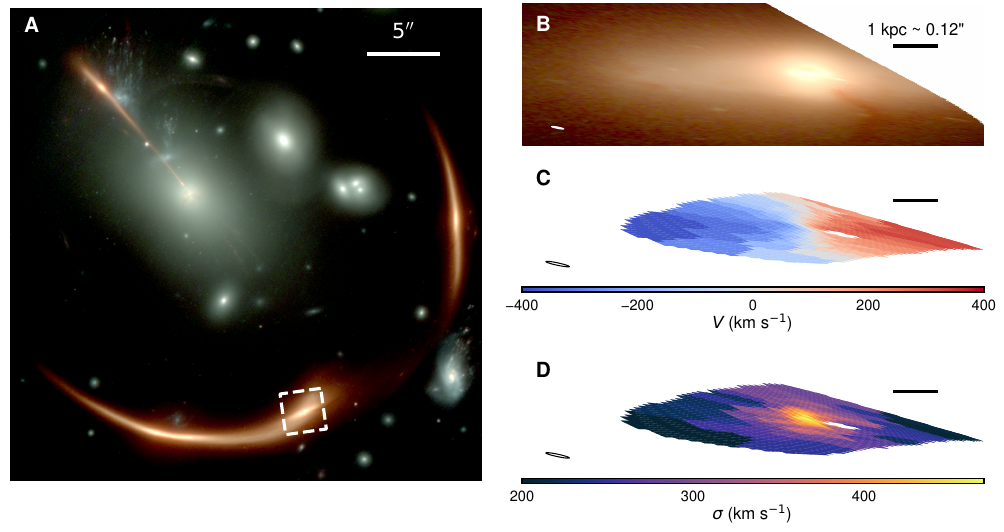}
    \caption{{\bf Morphology and and stellar kinematics of MRG-M0138.} \textbf{(A)} JWST/NIRCam image of the foreground galaxy cluster MACS~J0138.0--2155 through the F115W (blue), F150W (green), and F356W (red) filters, displayed with a logarithmic brightness stretch. The bright red arcs are multiple images of MRG-M0138 produced by gravitational lensing. The dashed box shows the field of view of the NIRSpec IFU used in our observations. \textbf{(B)} Reconstructed image of the galaxy in the source plane, after removing the effect of gravitational lensing. The white ellipse is the effective PSF. \textbf{(C)} Same region as panel B, colored to indicate the stellar velocity $V$ derived from the NIRSpec IFU data. \textbf{(D)} Same as panel C, but color indicates the stellar velocity dispersion $\sigma$.
    \label{fig1}}
\end{figure}

A rare class of galaxies in the local Universe, known as relics, have been interpreted as the descendants of galaxies that quenched at $z \gtrsim 2$ and experienced no significant mergers afterward, thereby surviving intact to $z \sim 0$ \cite{Trujillo14}. We test this interpretation by comparing the resolved stellar kinematics and $M_{\bullet}$ of MRG-M0138 to those of the relic galaxies. The rapidly rotating disk and peaked central velocity dispersion of MRG-M0138 (Fig.~\ref{fig1}C-D) resemble the kinematics of relic galaxies \cite{Yildirim15,FerreMateu17}. Stellar dynamical models indicate that the relic galaxies, like MRG-M0138, are outliers on the $M_{\bullet}$--$M_{\rm bulge}$ relation but not the $M_{\bullet}$--$\sigma$ relation \cite{Walsh-N1271,Walsh-N1277,Walsh-Mrk1216} (Figs.~\ref{fig3}C-D). Our results therefore support the interpretation of relic galaxies as being undisturbed descendants of early quiescent galaxies.



\begin{figure}
    \centering
    \includegraphics[width=\textwidth]{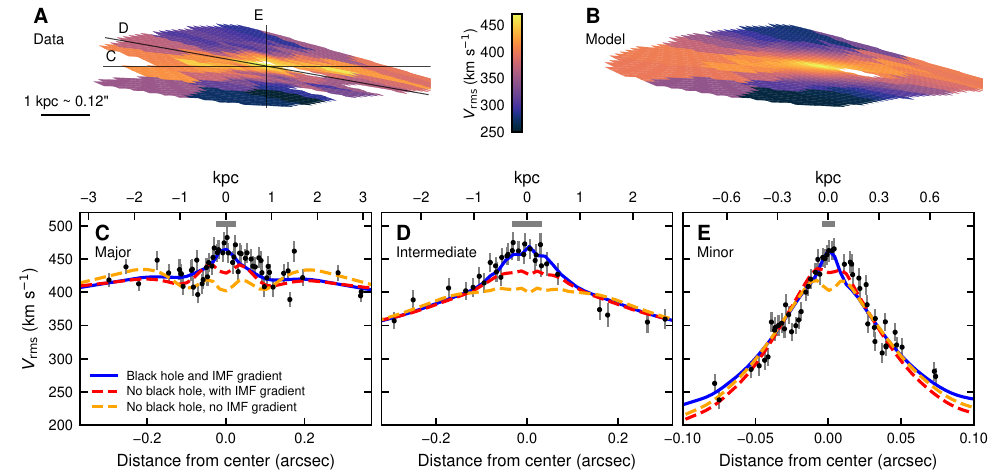} \vspace{2mm}
    \caption{{\bf The black hole mass constrained by stellar dynamical models.} \textbf{(A)} Same as Fig.~\ref{fig1}D, but showing the second velocity moment $V_{\rm rms}$ (color bar). Labeled lines indicate the directions along which kinematics are displayed in panels C-E. \textbf{(B)} The corresponding $V_{\rm rms}$ map produced by the best-fitting fiducial dynamical model, which includes a black hole. \textbf{(C)} The kinematics along the major axis of the galaxy. Colored lines indicate the fiducial model, which includes a black hole and a radial IMF gradient (blue solid line); a similar model that lacks a black hole (red dashed line); and a model that includes neither a black hole nor a radial IMF gradient (orange dashed line). These are respectively the models B-cyl, B-cyl-noBH, and A-cyl-noBH described in the supplementary material. The grey scale bar indicates the width of the PSF. Corrections were applied to the observed data to remove distortions caused by spatial binning \cite{methods}. \textbf{(D)} Same as panel C but showing kinematics along an intermediate axis, which makes a $45^{\circ}$ angle with the major and minor axes when deprojected. \textbf{(E)} Same as panel C but for the minor axis. The two models without a black hole do not match the central peak in $V_{\rm rms}$ (panels C-E).\label{fig2}}
\end{figure}

\begin{figure}
    \centering
    \includegraphics[width=0.75\textwidth]{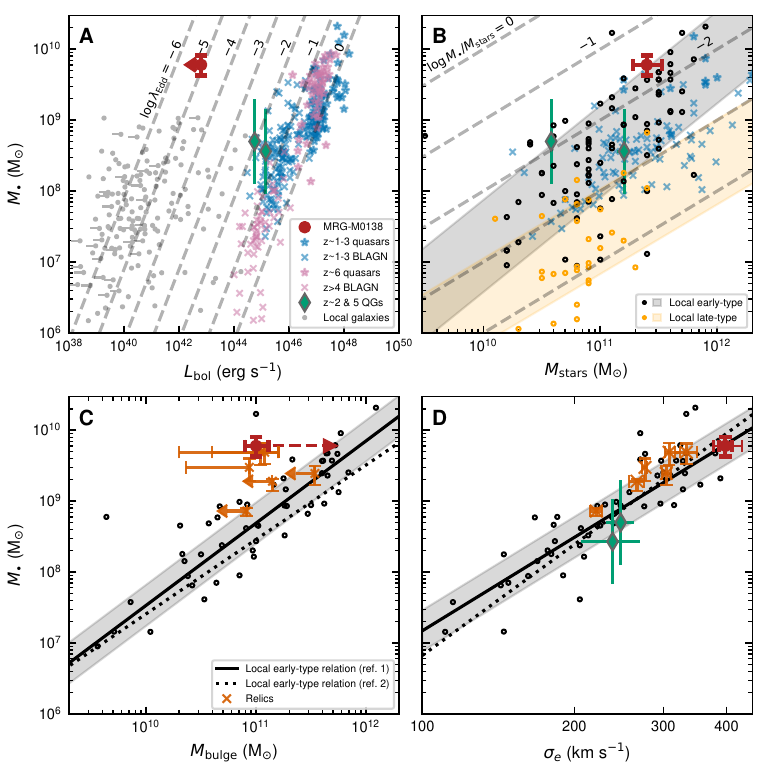}
    \vspace{-2ex}
    \caption{{\bf Comparison of MRG-M0138 to galaxy--black hole scaling relations.} Data sources are listed in Table~\ref{tab:fig3sources}. \textbf{(A)} Black hole mass as a function of AGN luminosity $L_{\rm bol}$. The grey dashed lines are logarithmically spaced values of the Eddington ratio. Our measurements of MRG-M0138 are shown with the red circle; the upper limit on $L_{\rm bol}$ includes the statistical $2\sigma$ range and systematic uncertainties. Other data points are observational measurements of quasars at $z \sim 1$ to 3 (blue stars) and $z \sim 6$ (pink stars), broad-line AGN (BLAGN) at $z \sim 1$ to 3 (blue crosses) and $z \gtrsim 4$ (pink crosses), two quiescent galaxies (QGs) at $z \sim 2$ and 5 with BLAGN (green diamonds), and a sample of local galaxies (grey dots, attached lines indicate $3\sigma$ upper limits). \textbf{(B)} The black hole mass as a function of total stellar mass $M_{\rm stars}$. Grey dashed lines show logarithmically spaced values of the mass ratio $M_{\bullet} / M_{\rm stars}$. The red circle is MRG-M0138, which is compared to local early-type (black circles) and late-type (orange circles) galaxies with dynamically determined $M_{\bullet}$ [with shaded bands showing the scaling relations previously fitted to those data sets \cite{Greene20}] and $z \gtrsim 1$ galaxies (symbols as in panel A). \textbf{(C)} The black hole mass as a function of bulge stellar mass $M_{\rm bulge}$. Black circles are local early-type galaxies, and black lines are  scaling relations previously derived from observations [solid line with shaded band showing the $1\sigma$ scatter \cite{KH13}, dashed line \cite{McConnell13}]. Oranges crosses are local relic galaxies, with error bars on $M_{\rm bulge}$ encompassing a range of morphological decompositions, and upper limits shown where the total stellar mass is plotted. The dashed red arrow indicates our estimated evolution of MRG-M0138 (see text). \textbf{(D)} Black hole mass as a function of effective stellar velocity dispersion $\sigma_e$. Symbols follow panels A and C. Thin error bars on MRG-M0138 encompass a range of $\sigma_e$ definitions \cite{methods}. Except where otherwise specified, error bars show $1\sigma$ uncertainties. Further details on the figure construction are provided in the supplementary materials.\label{fig3}}
\end{figure}


\clearpage 

%



\section*{Acknowledgments}

This work is based on observations made with the NASA/ESA/CSA James Webb Space Telescope. These observations are associated with programs GO-2345 and DD-6549. The data were obtained from the Mikulski Archive for Space Telescopes at the Space Telescope Science Institute, which is operated by the Association of Universities for Research in Astronomy, Inc., under NASA contract NAS 5-03127. This research has made use of data obtained from the Chandra Data Archive provided by the Chandra X-ray Center (CXC). A.B.N. performed part of this work at the Aspen Center for Physics, which is supported by National Science Foundation grant PHY-2210452.

\paragraph*{Funding:} A.B.N. and S.G. received support for program GO-2345 provided by NASA through a grant from the Space Telescope Science Institute. J.D.R.P. and A.B.N. received support for program DD-6549 provided by NASA through a grant from the Space Telescope Science Institute. A.Z. acknowledges support by Grant No. 2020750 from the United States--Israel Binational Science Foundation (BSF) and Grant No. 2109066 from the United States National Science Foundation (NSF); and by the Israel Science Foundation Grant No. 864/23. S.Bi. acknowledges funding from NASA ATP 80NSSC22K1897. R.S.E. acknowledges financial support from the Peter and Patricia Gruber Foundation. M.J.J. acknowledges support for the current research from the National Research Foundation (NRF) of Korea under the programs 2022R1A2C1003130 and RS-2023-00219959. S.S. received funding from the European Union’s Horizon 2022 research and innovation programme under the Marie Skłodowska-Curie grant agreement No 101105167 — FASTIDIoUS. M.O. was supported by JSPS KAKENHI Grant Numbers JP22K21349 and JP25H00662. Y.F. was supported by JSPS KAKENHI Grant Number JP23K13149. S.C. was supported by Basic Science Research Program through the National Research Foundation of Korea (NRF) funded by the Ministry of Education (No. RS-2024-00413036). A.A. acknowledges financial support through the Beatriz Galindo programme and the project PID2022-138896NB-C51 (MCIU/AEI/MINECO/FEDER, UE), Ministerio de Ciencia, Investigación y Universidades. S.E. and S.H.S. thank the Max Planck Society for support through the Max Planck Fellowship for S.H.S. S.H.S. received funding from the European Research Council (ERC) under the European Union's Horizon 2020 research and innovation programme (LENSNOVA: grant agreement No 771776). S.H.S. and G.C. were supported in part by the Deutsche Forschungsgemeinschaft (DFG, German Research Foundation) under Germany's Excellence Strategy -- EXC-2094 -- 390783311. G.G. acknowledges financial support through grant PRIN-MIUR 2020SKSTHZ. C.G., P.B., and P.R. acknowledge support through grant MIUR2020 SKSTHZ. J.E.G. acknowledges support from NSF AAG grant 2306950. S.Be. was supported by the ERC Starting Grant ‘Red Cardinal’, GA 101076080. J.L.W. acknowledges the support of NSF AST-2206219. K.G. and T.N. acknowledge support from Australian Research Council Laureate Fellowship FL180100060. K.G. and T.M.B. acknowledge support from Australian Research Council Discovery Project DP230101775.
\paragraph*{Author contributions:}
A.B.N. led the analysis of the source kinematics, stellar populations, and structure; the dynamical modeling; the X-ray analysis; and the drafting of the paper. M.G. reduced the spectroscopic data. A.M. searched for archival data. A.B.N., S.Be., R.S.E., K.G., M.J., M.K., T.N., A.M.K, and J.P. contributed to the acquisition, planning, and execution of the observations. S.Bi. and A.B.N. performed the Astrid comparison. S.H.S., S.E., G.C., G.G., C.G., S.S., A.A., P.B., S.C., J.M.D., N.F., B.F., Y.F., M.J.J., P.S.K., A.K.M., S.N., M.O., P.R., and A.Z. performed calculations based on seven lens models that enabled the source-plane reconstruction and estimation of its uncertainty. All authors contributed to the interpretation of the results and to the writing of the paper.

\paragraph*{Competing interests:}
There are no competing interests to declare.
\paragraph*{Data and materials availability:}
The JWST and HST observations are available from the Barbara A. Mikulski Archive for Space Telescopes (MAST) \cite{mastdataset}. Our derived stellar kinematics, the lens mapping, and custom analysis code are archived at Zenodo \cite{dataset}.


\subsection*{Supplementary materials}
Materials and Methods\\
Figures S1 to S10\\
Tables S1 to S3\\
References \textit{(56-120)}\\


\newpage


\renewcommand{\thefigure}{S\arabic{figure}}
\renewcommand{\thetable}{S\arabic{table}}
\renewcommand{\theequation}{S\arabic{equation}}
\renewcommand{\thepage}{S\arabic{page}}
\setcounter{figure}{0}
\setcounter{table}{0}
\setcounter{equation}{0}
\setcounter{page}{1} 


\begin{center}
\section*{Supplementary Materials for\\ \scititle}

Andrew~B.~Newman$^{\ast}$,
    Meng~Gu,
    Sirio~Belli,
    Richard~S.~Ellis,
    Sai Gangula,
    Jenny~E.~Greene,
    Jonelle~L.~Walsh,
    Sherry~H.~Suyu,
    Sebastian~Ertl,
    Gabriel~Caminha,
    Giovanni~Granata,
    Claudio~Grillo,
    Stefan~Schuldt,
    Tania~M.~Barone,
    Simeon~Bird,
    Karl~Glazebrook,
    Marziye~Jafariyazani,
    Mariska~Kriek,
    Allison~Matthews,
    Takahiro~Morishita,
    Themiya~Nanayakkara,
    Justin~D.~R.~Pierel,
    Ana~Acebr\'on,
    Pietro~Bergamini,
    Sangjun~Cha,
    Jose~M.~Diego,
    Nicholas~Foo,
    Brenda~Frye,
    Yoshinobu~Fudamoto,
    M.~James~Jee,
    Patrick~S.~Kamieneski,
    Anton~M.~Koekemoer,
    Asish~K.~Meena,
    Shun~Nishida,
    Masamune~Oguri,
    Piero~Rosati,
    Adi~Zitrin\\
\small$^\ast$Corresponding author. Email: anewman@carnegiescience.edu
\end{center}

\subsubsection*{This PDF file includes:}
Materials and Methods\\
Figures S1-S10\\
Tables S1-S3 \\
References \textit{(56-120)}\\

\newpage


\subsection*{Materials and Methods}

We used a flat $\Lambda$ cold dark matter cosmology with Hubble constant $H_0 = 67.74$~km~s$^{-1}$~Mpc$^{-1}$ and matter density $\Omega_m = 0.3075$ \cite{Planck15}. At the redshift of MRG-M0138, an angle of $1''$ in the source plane corresponds to 8.61~kpc. 

\subsubsection*{Observations}

Initial images of MRG-M0138 were obtained through the JWST program GO-2345. Following the discovery of a supernova in those images \cite{Pierel24}, deeper NIRCam images were obtained through program DD-6549, which provided 50 to 77~min of total exposure in each of the F115W, F150W, F200W, F277W, F356W, and F444W filters. We also used an 87~min observation from HST/ACS through the F555W filter (GO-14496).

The NIRSpec IFU observations (GO-2345) used the G140M/F100LP and G235M/F170LP grating/filter pairs. Because the source emission fills the IFU, we also observed off-source pointings to measure the background. The G235M observation on 2023 Nov 18 used the preset dither pattern {\tt 4-POINT-DITHER} to provide 127~min exposure in each of the on- and off-source fields. The G140M observation on 2023 Dec 27 provided 137~min exposure per field. 

\subsubsection*{Imaging Data Reduction}

We obtained uncalibrated NIRCam images from the MAST archive, processed by the JWST Science Data Processing system version 2023\_3a, and reduced them using the JWST Science Calibration Pipeline version 1.12.5 \cite{jwst1.12.5} and context file {\tt jwst\_1180.pmap}, with snowball detection activated. We used {\tt remstriping} \cite{Bagley23} to remove residual striping, masking out sources including the brightest cluster galaxy (BCG) of the lens. We obtained the reduced ACS image from the MAST archive and aligned it to the NIRCam images. All the images were resampled onto a 15~mas pixel scale using Gaia Data Release 3 \cite{GaiaDR3} as the astrometric reference. For each image, we built and subtracted a model of the BCG light using {\tt photutils.isophote} \cite{photutils}. We built empirical PSFs from isolated stars and constructed matching kernels using {\tt photutils.psf} \cite{photutils}.

\subsubsection*{NIRSpec Data Reduction}

The NIRSpec observations were reduced using the JWST Science Data Processing system version 2023\_3b and the JWST Science Calibration Pipeline version 1.15.1 \cite{jwst1.51.1}, with the context file {\tt jwst\_1276.pmap}. We activated the snowball rejection, {\tt NSClean}, and bad pixel self-care steps. We selected the exponential modified-Shepard method ({\tt emsm}) to resample the exposures onto an output frame with $0.05''$ spaxels (spatial pixels), with North upwards. We found that resampling onto a smaller spaxel scale slightly improves the spatial resolution. Although it also increases correlations among nearby spaxels and the amplitude of spectral wiggles (see below), we mitigate the wiggles and account for covariances in our dynamical modeling. We chose {\tt emsm} over the {\tt drizzle} resampling method, because we found that the latter method improved the effective angular resolution only slightly, at the cost of a large increase in the amplitude of the wiggles.

The on- and off-source observations were reduced separately and registered by adjusting the {\tt ra\_center} and {\tt dec\_center} of the off-source data cubes to pixel-align them with the corresponding on-source cubes, as required for background subtraction. The G140M and G235M on-source observations were registered to sub-spaxel precision by making slight adjustments to {\tt ra\_center} and {\tt dec\_center} until a map showing the ratio of the G140M and G235M flux densities, evaluated as the median over the wavelength interval common to both gratings, was featureless. We masked the edge area that did not receive full exposure with both gratings, resulting in higher noise and poorer outlier rejection. (An instrument problem required the second NIRSpec observation to be rescheduled and observed at PA rotated by $16^{\circ}$ from the first.) The remaining area was 8.0 arcsec${}^2$. 

We cleaned the off-source data cubes of outlier pixels and those with elevated noise. Because the background is mostly spatially uniform, we produced a local background estimate by applying a $3\times3$ spatial median filter to the cleaned off-source data cube, which reduced random noise and outliers. There were, however, some regions where the background intensity was locally elevated due to instrumental effects (e.g., stuck shutters in the micro-shutter assembly). The background in such regions was instead removed by direct subtraction: specifically, in spaxels where the background level differed by $> 10\%$ from the median, we estimated the local background spectrum directly from the cleaned off-source data cube without applying any smoothing. Finally we subtracted this background model from the on-source data cube and propagated the uncertainties.

\subsubsection*{Mitigating spectral wiggles}

The NIRSpec detectors undersample the PSF, and the images were resampled during data reduction, which imprints sinusoidal fluctuations (wiggles) on the spectra, particularly in regions with strong flux gradients \cite{Law23}. We developed a Fourier method to model and mitigate the wiggles. Due to flux conservation, wiggles become weaker as the flux is integrated over progressively larger apertures. We therefore examined the ratio between the spectrum in a given spaxel and the mean spectrum in a $5 \times 5$ neighborhood ($0.25'' \times 0.25''$) centered on that spaxel. Because MRG-M0138 has a smooth light distribution and is highly resolved, the differences between spectra separated by a few spaxels are small, and therefore the ratio spectrum is dominated by any wiggles.

\begin{figure} 
	\includegraphics[width=\linewidth]{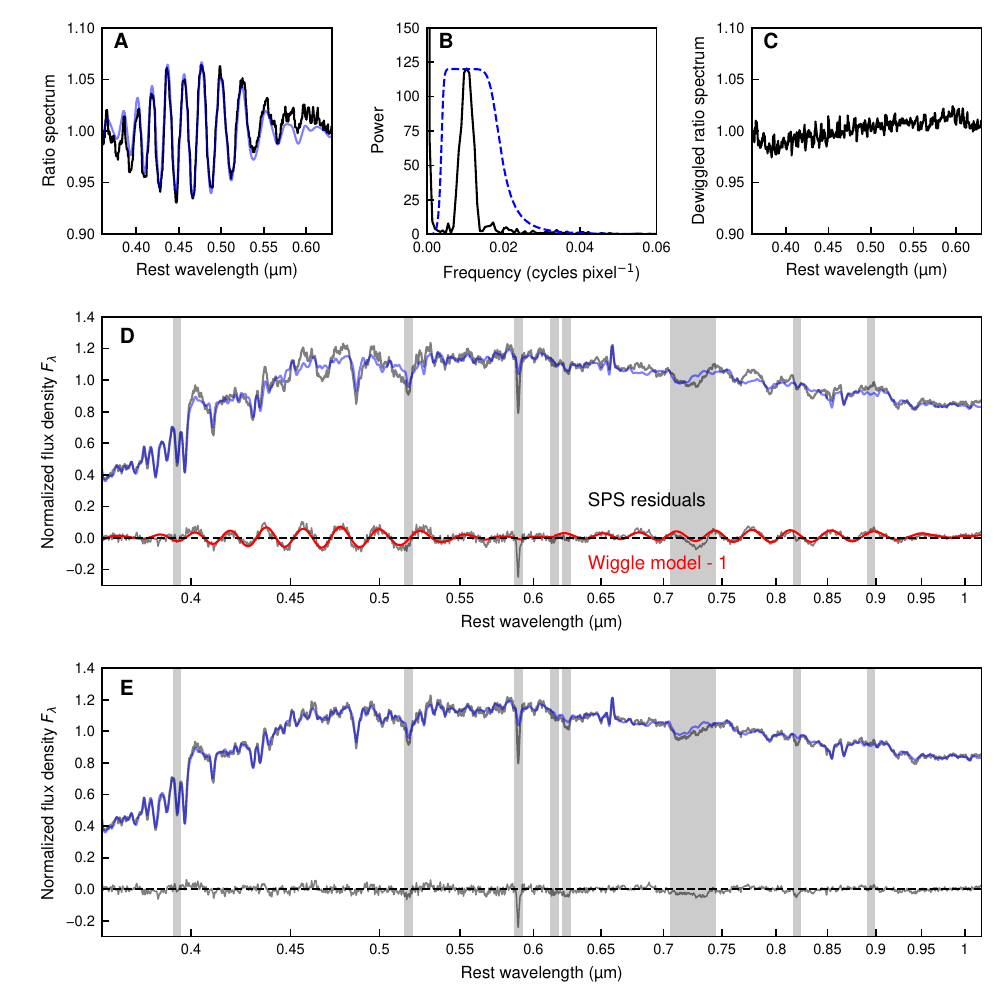}
	\caption{\textbf{Modeling of spectral wiggles induced by NIRSpec undersampling.} {\bf (A)} An example ratio spectrum (black curve), formed by taking the ratio of the G140M spectrum in an example spaxel to the mean of its neighborhood, and the associated wiggle model (blue) after filtering. {\bf (B)} The power spectrum of the ratio spectrum from panel A (black line) and the Butterworth filter response (blue dashed line, arbitrary scaling). {\bf (C)} The de-wiggled ratio spectrum, formed by dividing the ratio spectrum and the wiggle model shown in panel A. {\bf (D)} The full spectrum in this example spaxel, without wiggle correction (grey line), and an SPS model fitted to the wiggle-corrected spectrum (blue line). Residuals (black line) are compared to the wiggle model (red line). The dashed line is the zero level, and the grey shaded regions were masked out (see text). {\bf (E)} Same as panel D but after application of the wiggle correction.}
	\label{fig:dewiggling} 
\end{figure}

Fig.~\ref{fig:dewiggling}A shows an example of a ratio spectrum, with light smoothing applied. There are quasi-periodic modulations, with their power spectrum shown in Fig.~\ref{fig:dewiggling}B. We aimed to isolate signal at the frequencies characteristic of wiggles while preserving signal at higher frequencies (which can arise from noise or local variations in absorption lines) and lower frequencies (which can arise from local color variations). We used a fifth-order digital Butterworth bandpass filter ({\tt scipy.signal.butter}, \cite{scipy}). By examining many spaxels, we estimated critical frequencies of 0.004 and 0.018 cycles pixel${}^{-1}$ (half the {\tt Wn} parameter used by {\tt scipy.signal.butter}). This frequency range corresponds to 6 to 28 µm${}^{-1}$ and 4 to 17 µm$^{-1}$ for the G140M and G235M gratings, respectively. In high-resolution grating observations, other studies have found wiggle frequencies in the range 5 to 60 µm$^{-1}$ around 2.4~µm \cite{Perna23,Dumont25}; considering the higher dispersion, we would then expect to find a range of about 2 to 22 µm$^{-1}$ for G235M, which is similar to our findings. Fig.~\ref{fig:dewiggling}B shows that the bandpass filter selects the frequency range characteristic of the wiggles.

Fig.~\ref{fig:dewiggling}A also shows the result of filtering the ratio spectrum, producing a wiggle model. Dividing the ratio spectrum by the wiggle model cleans it of wiggles while preserving the noise characteristics and broadband features (Fig.~\ref{fig:dewiggling}C). Figs.~\ref{fig:dewiggling}D and E show the full uncorrected and corrected spectra, respectively; the corrected spectrum is formed by dividing the uncorrected one by the wiggle model. Both panels also compare to an SPS model (see below) fitted to the corrected spectrum. Residuals between this model and the uncorrected spectrum are large and strongly correlated with the wiggle model (Fig.~\ref{fig:dewiggling}D), while the residuals between the model and the corrected spectrum are small (Fig.~\ref{fig:dewiggling}E).

We consider our procedure effective in spaxels with ${\rm SNR} > 30$ and applied the wiggle correction to all such spaxels. At lower SNR, the amplitude of the modeled wiggles increased rapidly. Because the lower-SNR spaxels are farther from the galaxy center, where flux gradients are generally smaller and therefore weaker wiggles are expected, the higher observed amplitudes indicate that the estimated correction is dominated by noise, so we did not apply it.

\subsubsection*{Flux calibration corrections}

To test the relative flux calibration of the NIRSpec spectra, we produced an integrated spectrum of MRG-M0138 and fitted it with an SPS model (see below). Due to the wide aperture, wiggles are negligible in this case. We found that the continuum shape of the observations and the best-fitting model are similar but exhibit small systematic residuals. We brought the data and models into better agreement as follows: we divided the integrated spectrum by the best-fitting model, fitted the result with a 15th order polynomial, then divided this polynomial from the spectrum in every spaxel of the data cube. The flux corrections are $\lesssim 2\%$. Because they affect only velocity scales $\gtrsim 2 \times 10^4$~\kms, they have no direct influence on the kinematics. 

\subsubsection*{NIRSpec PSF}

Determining the PSF of the NIRSpec IFU observations is not trivial. We used methods similar to prevous work \cite{DEugenio24} to derive the effective PSF in two ways. 

\emph{PSF A}: We estimated the PSF from observations of the standard star GSPC P330-E (JWST program CAL-1538). The data were reduced using the same resampling parameters as applied to MRG-M0138. The {\tt 4-POINT-NOD} pattern used in those observations differs from those of MRG-M0138, but it samples the same subpixel phases. For each of the G140M and G235M gratings, we constructed a PSF by taking the median over the wavelength range used in our SPS modeling ($\lambda \approx 1.1$ to 3.0~µm). We then averaged the G140M and G235M PSFs and modeled the result as the sum of three anisotropic Gaussian components. The core component contains 81\% of the light and has a circularized full width at half maximum (FWHM) of $0.161''$. The core is elongated along the direction of the slices (in our model, within $2^{\circ}$), as also found in prevoius work \cite{DEugenio24}. To apply the fitted PSF model to analyses of our data cube, we rotated it to maintain its orientation with respect to the instrument axes. Fitting the G140M or G235M PSFs individually changed the FWHM by $\lesssim 10\%$, indicating that the size of the effective PSF (including pixel convolution) depends weakly on wavelength below 3~µm \cite{DEugenio24}.

\emph{PSF B}: We estimated the NIRSpec PSF in situ by degrading the NIRCam resolution. We rebinned the NIRCam F200W image to match the IFU data cube and synthesized F200W photometry from the spectra. We then found the six-parameter anisotropic Gaussian kernel (amplitude, center $x_0$, $y_0$, dispersions $\sigma_x$, $\sigma_y$, and $\theta$) that, when convolved with the NIRCam image, best matches the synthetic NIRSpec image. The NIRSpec PSF is then the convolution of this kernel with the NIRCam PSF. Because this method does not have the flexibility to match the broad PSF components, we considered only the core of the NIRCam PSF, which was approximated as an isotropic Gaussian with a FWHM of $0.066''$. Convolving this with the best-fitting kernel results in an estimated FWHM of $0.127''$ for the NIRSpec PSF core, which is 21\% smaller than the core of PSF~A. Previous work \cite{DEugenio24} also found that a similar method produced a more compact PSF than a standard star by a similar ratio. To produce PSF~B, we isotropically rescaled the core component of PSF~A to a circularized FWHM of $0.127''$, while maintaining the broader components.

We adopt PSF~B in our subsequent analysis, and we test the effect of instead using PSF~A in the dynamical models.

\subsubsection*{Spectral resolution}

Our kinematic measurements adopt the spectral resolution for slice-filling emission from the JWST documentation \cite{JDox}. The effective spectral resolution for a given data cube, however, can be affected by the resampling method \cite{Law23}, which we tested using archival observations of the Ring Nebula with the NIRSpec IFU (GO-1558). We used this data set because i) observations were made with the same dither pattern, ii) we reduced them using the same {\tt emsm} resampling method, and iii) the extended emission approximately uniformly illuminates the slices, as for MRG-M0138, and iv) the lines can be are spectrally unresolved, because the Ring Nebula expansion velocity of 30~\kms~\cite{Kastner25} is much smaller than the instrumental resolution $\sigma_{\rm inst} \approx 130$~\kms. We fitted Gaussian profiles to numerous unblended emission lines in this data cube to measure the effective spectral resolution as a function of wavelength for the G140M and G235M gratings. We found that it agrees with the documentation \cite{JDox} within 3\%. Because the stellar velocity dispersions $\sigma$ are always resolved ($\sigma > \sigma_{\rm inst}$), a 3\% error in the resolution would have a negligible effect on the measured $\sigma$ ($\sim$1~\kms~in the galaxy center and 3~\kms~in the outskirts).

\subsubsection*{Spatial binning and preparation for modeling}

We logarithmically rebinned the G140M and G235M spectra and their variances onto a common wavelength grid with 132~\kms~per pixel. The two data cubes were shifted to the rest frame, using a systemic redshift $z_{\rm sys} = 1.9480$ determined from initial dynamical models (see below), and spliced together at a transition wavelength of $\lambda_{\rm rest} = 0.615$~µm.  We corrected the spectra and images for foreground Milky Way extinction \cite{SF11}. SPS models (see below) were fitted to the spectra in each individual spaxel. These initial fits were only used to  identify outlier pixels ($>3\sigma$, 0.6\% of pixels) in an iterative process, and to adjust the pipeline noise spectra \cite{Cappellari23}, which we rescaled by factors of $\simeq 1.2$ to 2.0 to achieve a reduced $\chi^2_{\nu} = 1$. Noise rescaling by similar factors is common \cite{Maseda23}. We defined the signal-to-noise ratio (SNR) of a spectrum as the median ratio using these rescaled noise estimates.

We spatially binned the spectra to a target SNR of 40 using the software {\tt vorbin} version 3.1.5 \cite{Cappellari03,vorbin}. The binning was applied to the 1341 spaxels with a F200W surface brightness exceeding 1.0 megaJansky~(MJy)~sr${}^{-1}$, excluding 10 spaxels contaminated by the supernova. To account for correlated noise in nearby spaxels, which is substantial due to the resampling, we provided {\tt vorbin} with a custom function to fit SPS models to a binned spectrum, from which we estimated the noise and identified outliers as described above. We arrived at 219 Voronoi bins with ${\rm SNR} = 23$ to 64 (median 42; see Fig.~\ref{fig:image_plane_bins}) and sizes of 1 to 68 spaxels, with 129 of the bins being single spaxels. Within each bin, we derived matched photometry from the average surface brightness of the constituent pixels in each of the seven images, convolved to match the broadest PSF (F444W). Its FWHM is $0.147''$, which is close to the NIRSpec IFU PSF. Because the uncertainties in the photometry are dominated by small systematics, we added in quadrature 3\% of the maximum flux density $F_{\lambda}$ (which was always in F150W) in each spatial bin, similar to previous work \cite{Cappellari23}.

\subsubsection*{Stellar population synthesis modeling}

We used {\tt ppxf} version 9.1.1 \cite{Cappellari04,Cappellari20,Cappellari23,ppxf} to model the spectrophotometry and measure stellar populations and kinematics. The input SPS models were simple stellar populations from the E-MILES library \cite{Vazdekis16}, which {\tt ppxf} combined with regularized weights to allow for flexible star formation histories. To cover the necessary wavelength range, we restricted the simple stellar populations to ages of 0.14 to 4~Gyr and metallicities ${\rm [Z/H]} = -0.4$, 0, and 0.22. These limits were appropriate because previous work has shown that MRG-M0138 is metal-rich \cite{Jafariyazani20} and quiescent \cite{Newman18a}. The upper limit on age permitted stars slightly older than the Universe, to allow for model uncertainties. The models were built with a Salpeter IMF, but we rescaled all stellar masses by a factor 0.66 so that the SPS-based mass estimates $M_{*, {\rm MW}}$ are equivalent to a Milky Way IMF \cite{Kroupa01}. (This choice of fiducial IMF does not affect the dynamical modeling, because the SPS-based stellar masses are later rescaled to fit the kinematics.)  We adopted a dust attenuation law measured in local galaxies \cite{Calzetti00}. The model spectrum also includes emission lines emission from the hydrogen Balmer series, [O~\textsc{ii}] $\lambda\lambda$3727,3730, [O~\textsc{iii}] $\lambda\lambda$4960,5008, [O~\textsc{i}] $\lambda\lambda$6302,6366, [N~\textsc{ii}] $\lambda\lambda$6550,6585, and [S~\textsc{ii}] $\lambda\lambda$6718,6733, where the $\lambda\lambda$ notation indicates the wavelengths of doublet components in \AA ngstroms. Because the Balmer emission lines were not detected, we assumed the emission lines have the same attenuation as the starlight. The model spectra were convolved to match the wavelength-dependent instrumental resolution.

We masked $\pm 1500$~\kms~intervals around several features: Na~\textsc{i}~D and Ca~\textsc{ii}~K, which may be affected by gas absorption; Mg~\textsc{i}~b, Na~\textsc{i}~8190, and TiO2 \cite{Spiniello14}, which are affected by $\alpha$-element enhancement and the IMF and showed residuals in some spectra; and transitions at 6150~\AA~(G140M to G235M) and 8949~\AA~(resolution change in E-MILES). We also masked $\pm$200~\AA~around 7250~\AA~due to systematic residuals.

\begin{figure} 
	\centering
	\includegraphics[width=5in]{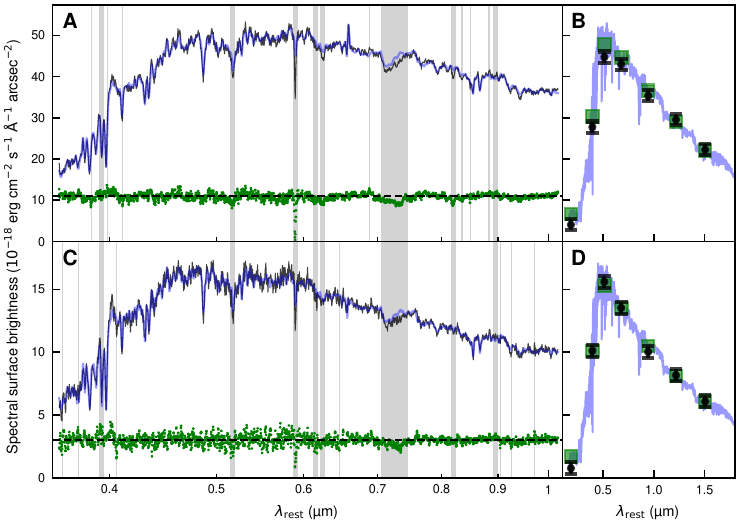}
	\caption{\textbf{Spectrophotometry and SPS model fits in two spatial bins.} \textbf{(A)} The observed spectrum (black line) and fitted model (blue line) for the central spaxel. Residuals are shown as green points, with the dashed line indicating the shifted zero point. Grey bands indicate masked spectral pixels: thin bands are outlier pixels rejected during the fitting process, and thick bands are wavelength regions we chose to mask out (see text). \textbf{(B)} The observed photometry (black circles with $1\sigma$ uncertainties), model spectrum (blue line), and synthesized model photometry (green boxes). \textbf{(C-D)} Same as panels A-B, but for the spatial bin with the median SNR.
    \label{fig:example_spec_fits}}
\end{figure}

To map the stellar mass-to-light ratio $(M_*/L)_{\rm MW}$, which we use to model the stellar mass distribution (see below), we simultaneously fitted the seven-band photometry and the spectra from $\lambda_{\rm rest} = 0.36$ to 1.02~µm. We allowed the models to be modulated by a multiplicative polynomial with order $N_{\rm mult} = 3$ to mitigate flux calibration errors. In each spatial bin, we measured $(M_* / L)_{\rm MW}$ in the F200W filter shifted by a factor $(1 + z_{\rm sys})^{-1}$ (to $\lambda_{\rm rest} \approx 0.68$~µm). Fig.~\ref{fig:example_spec_fits} compares the resulting data and models.

\subsubsection*{Stellar kinematics}

\begin{figure} 
	\centering
	\includegraphics[width=5in]{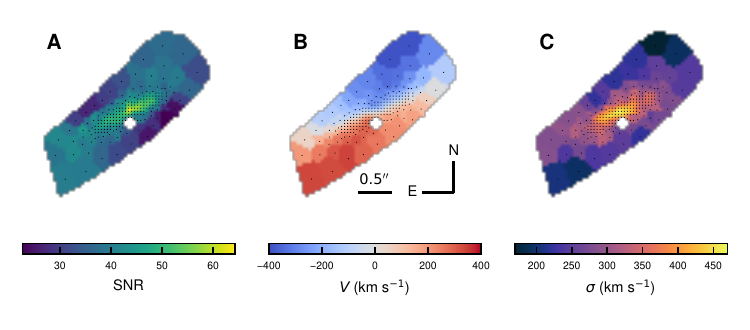}
	\caption{\textbf{Voronoi binning and image plane stellar kinematics.} \textbf{(A)} The SNR of the Voronoi binned spectra. Points show the center of each bin. The white cross south of the center was excluded from the binning due to contamination from a supernova. \textbf{(B)} The stellar velocity in each bin. \textbf{(C)} The stellar velocity dispersion in each bin.
    \label{fig:image_plane_bins}}
\end{figure}

We performed separate {\tt ppxf} fitting to measure the stellar $V$ and $\sigma$ in each binned spectrum. This followed the process described above for our $(M_*/L)_{\rm MW}$ measurements, but with the following differences: we omitted the wavelength range $<0.40$~µm, which reduced the sensitivity of the kinematics to the polynomial orders; we included an additive polynomial with order $N_{\rm add} = 2$, which reduces template mismatch; and we omitted the photometry so did not model dust attenuation, which was subsumed into the multiplicative polynomial. Fig.~\ref{fig:image_plane_bins} shows the kinematic maps in the image plane, and Fig.~\ref{fig1} shows the source plane.

We estimated the statistical uncertainties in $V$ and $\sigma$ by comparing measurements derived from the G140M and G235M spectra separately. We computed a reduced $\chi^2_{\nu} = 1.9$ to 2.1 for the G140M--G235M differences in $V$ and $\sigma$, using the uncertainties returned by {\tt ppxf}. We therefore increased these by a factor of 1.4, resulting in median $1\sigma$ uncertainties in $V$ and $\sigma$ of 10~\kms~and 13~\kms, respectively. 

Systematic uncertainties in $\sigma$ were evaluated by varying several analysis choices and measuring changes to the median $\sigma$ over the 30 highest-SNR bins. First, we varied the polynomial orders over $0 \leq N_{\rm mult} \leq 6$ and $-1 \leq N_{\rm add} \leq 6$, where $-1$ indicates no polynomial. We found the results are stable to these choices: provided that $N_{\rm mult} \geq 2$ and $N_{\rm add} \geq 0$, systematic changes to $\sigma$ were $\lesssim 1$\%. Second, we extracted kinematics from narrow 0.1~µm-wide chunks to evaluate the sensitivity to wavelength. For this test we fixed the template to the one determined from the full spectral range, and we used low-order polynomials ($N_{\rm mult} = 1$ and $N_{\rm add} = -1$) given the narrow wavelength range. Except in the 0.7 to 0.8~µm interval, which is mostly masked, we found that the median $\sigma$ varied by $\pm 2.7$\% (rms). Third, we found that $\sigma$ measurements derived using the stellar population settings in the previous section were  lower by 2\%. Fourth, we extracted kinematics using {\tt FSPS} templates \cite{Conroy09,Conroy10} and found no systematic shift. Altogether we estimated a systematic uncertainty of 3\% in $\sigma$. 

\begin{figure} 
	\centering
	\includegraphics[width=5in]{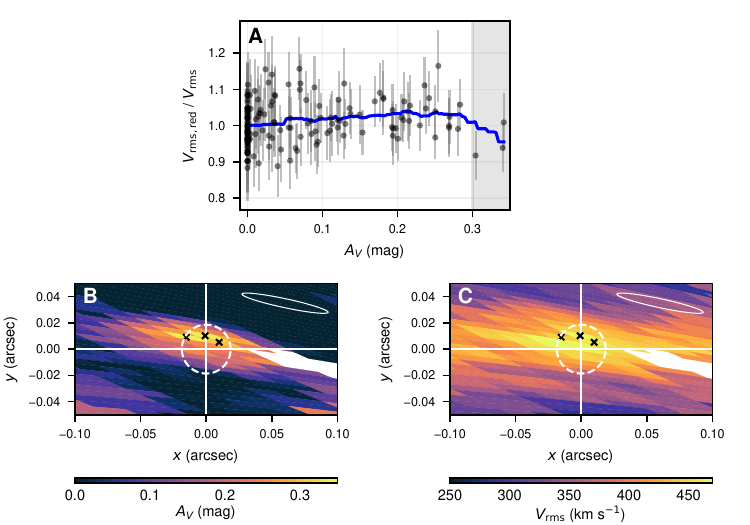}
	\caption{\textbf{Tests of the influence of dust on the inner stellar kinematics.} \textbf{(A)} The ratio of $V_{\rm rms, \rm red}$, the second velocity moment measured using only the red portion of the spectrum ($\lambda_{\rm rest} > 0.8$~µm), to the fiducial value $V_{\rm rms}$, as a function of $A_V$ for spaxels within $0.1''$ (source plane) of the galaxy center. The blue line is a running mean. We find systematic wavelength-dependent differences only in the three spaxels with $A_V > 0.3$~mag. \textbf{(B)} Source-plane map of $A_V$ in the inner region as derived from SPS modeling. Black crosses show the bins masked in the dynamical modeling due to high $A_V$. The PSF is shown by the white ellipse in the upper right corner. The radius of the white dashed circle is $r_{\rm inf} = 19$~mas. $x$ and $y$ are coordinates that are aligned with major and minor axes, respectively, and have their origin at the galaxy center. \textbf{(C)} Same as panel B but mapping $V_{\rm rms}$.
    \label{fig:dust}}
\end{figure}

MRG-M0138 is globally dust-poor \cite{Whitaker21}, yet our SPS modeling indicates some regions with modest attenuation, including near the center (Fig.~\ref{fig:dust}B). To test whether dust affects the observed line-of-sight kinematics, we compared our fiducial kinematic measures to a second set derived by fitting only the reddest part of the spectrum at $\lambda_{\rm rest} > 0.8$~µm, including the calcium triplet. In the inner galaxy, we found no systematic differences or increased scatter between the two $V_{\rm rms}$ measures for the vast majority of bins with $V$-band attenuation $A_V < 0.3$~mag (Fig.~\ref{fig:dust}). This indicates that dust generally has a minimal effect on the observed kinematics. The three inner bins with $A_V = 0.3$ to 0.34~mag (Fig.~\ref{fig:dust}B-C) do show possible dust effects, and we therefore masked them for our dynamical modeling, along with two additional bins with $A_V > 0.3$~mag at larger radius.

We tested whether the wiggle correction biases the stellar kinematics. We do not expect a significant bias, because i) we Fourier filter ratios of similar spectra, not the spectra themselves, and ii) the critical frequencies correspond to large velocity scales of 0.7 to $3.3 \times 10^4$~\kms~cycle${}^{-1}$. As a test, we averaged the spectra within a $9 \times 9$ spaxel aperture centered on the galaxy nucleus in the uncorrected data cube (excluding spaxels near the supernova). Such an aperture is large enough that we expect the wiggles to average down to $\lesssim 1$\% \cite{Law23}. We extracted kinematics from this spectrum and computed $V_{\rm rms}$. This was compared to the value obtained from the wiggle-corrected data cube, by fitting the individual spaxels within the same aperture and computing their intensity-weighted $V_{\rm rms}$. The two $V_{\rm rms}$ measures differ by 1.9\%, which is smaller than the systematic uncertainty estimated above.

\subsubsection*{Lens models}

We adopt a lens model from previous work \cite{Ertl25}, which was constrained by the positions of 23 images of 8 sources with 4 different spectroscopic redshifts. Specifically, we used the most probable parameters from their {\tt iso\_halo} model, which we refer to as the fiducial lens model. We generated maps of the deflection angle on the pixel grids of our images and the IFU data cube, enabling the source-plane reconstructions and dynamical modeling.

Systematic uncertainties, which are typically dominant, were evaluated by comparing results from seven lens modeling groups who employed a wide variety of modeling methods. Estimates of the magnification of a supernova close to the center MRG-M0138 have previously been compared; for our NIRSpec-targeted multiple image, the magnification was found to vary by $-39$\% to +44\% from the fiducial model [\cite{Suyu25}, their ultimate models based on gold multiple images]. We evaluated the magnification at the center of MRG-M0138 using the fiducial model and adopted the same fractional uncertainties as at the position of the nearby supernova, finding $\mu = 29^{+13}_{-11}$. Earlier lens models indicated lower magnifications \cite{Newman18a,Rodney21}, but they were based on many fewer constraints. 

The areal magnification $\mu$ affects the inferred luminosity and mass of MRG-M0138, but the magnification is anisotropic. We evaluated the uncertainty in the ratio of the tangential to radial magnification $\mu_t / \mu_r$, which affects the inferred galaxy ellipticity. We found that the models agree much more closely on $\mu_t / \mu_r$ than on $\mu$, so the uncertainty in $\mu_t / \mu_r$ contributes negligibly to the systematic uncertainty in $M_{\bullet}$.

We determined the spatial resolution of the observations at the center of MRG-M0138 by casting PSFs to the source plane using the fiducial lens model and fitting an anisotropic 2D Gaussian. The PSFs have source-plane sizes, expressed in terms of the standard deviation, of $1.9 \times 17$ milliarcseconds (mas) (equivalent to 16 $\times$ 149~pc) for the NIRCam F200W image and 3.4~mas $\times$ 34~mas (29 $\times$ 288~pc) for the NIRSpec IFU data, both elongated $12^{\circ}$ from the major axis. The effective resolution $\sigma_{\rm psf}$ is the geometric mean of the two dimensions: $\sigma_{\rm psf} = 5.7$~mas (F200W), equivalent to 49~pc, and $\sigma_{\rm psf} = 11$~mas (IFU), equivalent to 91~pc.

\subsubsection*{Stellar light and mass distributions}

\begin{figure} 
	\centering
	\includegraphics[width=5in]{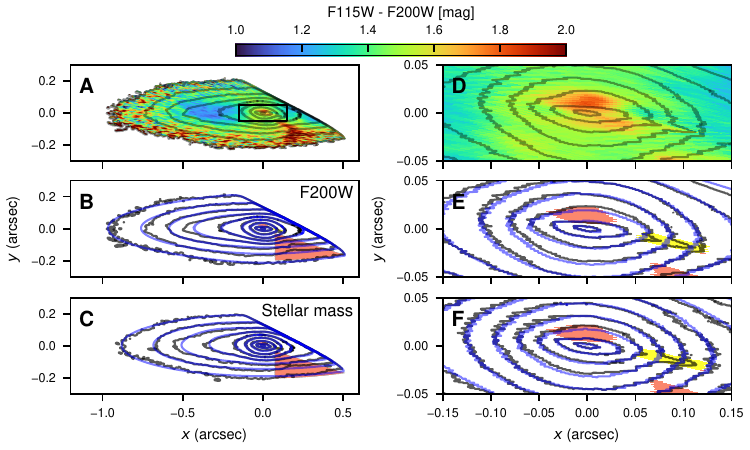}
	\caption{\textbf{Models of the stellar light and mass distributions}. \textbf{(A)} Source-plane map of the F115W--F200W color. Grey contours are F200W isophotes, spaced by 0.3 dex in surface brightness. \textbf{(B)} Observed F200W isophotes (grey contours), as in panel A, and the MGE model isophotes (blue). The red region is masked due to dust. \textbf{(C)} Same as panel B, but showing the stellar mass surface density, as estimated using the color--$M_*/L$ relation derived from the SPS models. \textbf{(D-F)} Same as panels A-C, but zoomed into the central region enclosed by the rectangle in panel A, and with isophotes spaced by 0.2~dex. The yellow regions in panels E and F were masked due to the supernova. The $x$ and $y$ coordinates are the same as in Fig.~\ref{fig:dust}.
    \label{fig:structure}}
\end{figure}

We modeled the distributions of stellar light and mass using multi-Gaussian expansions (MGEs) \cite{Emsellem11}, as required for JAM. The light map was computed by multiplying the F200W image, which probes wavelengths near the middle of the range that was used to extract kinematics, by a factor 4802 ${\rm L}_{\odot}$~pc${}^{-2}$ (MJy~sr${}^{-1}$)${}^{-1}$. Here ${\rm L}_{\odot}$ represents the spectral luminosity density of the Sun averaged over the F200W filter bandpass, shifted by a factor $(1 + z_{\rm sys})^{-1}$. We defined the center as the surface brightness peak, after light smoothing, and found it to exhibit a dispersion of 1~mas (source plane) among the NIRCam images, indicating it is not affected by dust. To identify dusty regions to exclude from the model fitting, we constructed a F115W--F200W color map (Fig.~\ref{fig:structure}A and D). By examining this map and residuals from initial MGE models, we identified two dusty regions located above the center and in the lower-right part of the disk. We also masked the vicinity of the supernova (Fig.~\ref{fig:structure}).

We fitted an initial model to the F200W image, resampled to the source plane, using {\tt mgefit} version 5.0.15 \cite{Cappellari02,mgefit}. We used the bulge-disk mode, which  simplifies the treatment of possible IMF gradients in our dynamical models. In this mode, each Gaussian component has one of two axial ratios, $q_{\rm disk}$ or $q_{\rm bulge}$. We then refined this initial {\tt mgefit} model by fitting in the image plane with a custom code, to properly account for the PSF. The model parameters were $\sigma_1$, ..., $\sigma_N$, $q_{\rm bulge}$, $q_{\rm disk}$, and position angle (PA). $N$ represents the number of MGE components with {\tt mgefit} $\sigma_i < 0.1''$. The other $\sigma$'s were fixed to their {\tt mgefit} values, because we expect them to be minimally affected by the PSF. As in {\tt mgefit}, we separated these non-linear parameters from the normalizations of the MGE components, which we instead determined using linear methods. To speed convergence, we adopted Gaussian priors centered on the {\tt mgefit} values, except for the size of the innermost component $\sigma_1$. For that parameter, we used a uniform prior with a lower bound $\sigma_{\rm min}$. Because the semi-minor axis of the NIRCam F200W PSF in the source plane has a size of 1.9~mas and is oriented close to the galaxy minor axis, the smallest resolvable isophote has a semi-major axis of $0.75 \times 1.9$~mas~/~$q$, where $q$ is the axial ratio and the factor 0.75 allows for structure slightly smaller than the PSF, as in {\tt mgefit}. Taking the smaller axial ratio ($q_{\rm disk}$) then yields $\sigma_{\rm min} \approx 7$~mas. For each considered parameter set, we generated the image of each MGE component traced through the lens mapping, convolved these by the PSF, solved for the normalizations, and computed the likelihood. We used a Gaussian likelihood with errors proportional to $(F_j R_j)^{-1}$, where $R_j$ is an elliptical radius and $F_j$ is the surface brightness of the {\tt mgefit} model at pixel $j$, which approximates the treatment by {\tt mgefit} of equal fractional errors in logarithmic bins of radius. To avoid diverging weights in the noisy galaxy outskirts, we enforced a floor on $F_j$ equal to 1/300 of the maximum $F_j$. Finally we used {\tt UltraNest} version 3.6.4 \cite{UltraNest} to find the maximum likelihood model. 

Figs.~\ref{fig:structure}B and E compare the resulting MGE to the observed image. The apparent isophote twist near the center is produced by the anisotropy of the PSF in the source plane and is not intrinsic to the model, which is axisymmetric. The largest residuals are close to the major axis, $\sim0.4''$ left of the center (Fig.~\ref{fig:structure}B). The excess light here seems to reflect a real asymmetry in part of the outer disk. In our dynamical models, we also found elevated residuals in this region and therefore masked two spatial bins. 

Dynamical modeling also requires the distribution of stellar mass. Our SPS modeling provided an estimate of $(M_*/L)_{\rm MW}$ for each binned spectrum. To transfer these estimates onto the higher-resolution NIRCam images, we fitted a linear relation $(M_* / L)_{\rm MW} / {\rm (M_{\odot}/L_{\odot})} = 0.80 \times ({\rm F115W} - {\rm F200W}) - 0.26$. Applying this relation to each spatial bin produces a range $(M_* / L)_{\rm MW} \approx 0.7$ to 1.1~${\rm M}_{\odot}/{\rm L}_{\odot}$. We used this relation to construct a map of the stellar mass surface density from the F115W and F200W images, which we then fitted with an MGE model using the same methods described above (Fig.~\ref{fig:structure}C and F). We used rest-frame optical images rather than rest-frame near-infrared images (e.g., F444W) to trace the stellar mass, because they have superior resolution and are less affected by stellar evolution uncertainties \cite{Maraston06}. The MGE components of the light and stellar mass models are listed in Table~\ref{tab:mge}. This procedure provided our fiducial stellar mass distribution, from which our dynamical modeling below allows large deviations.

\subsubsection*{Galaxy structural and kinematic properties}

$R_e$ measurements were derived from the F200W MGE model using the {\tt mge\_half\_light\_isophote} routine \cite{mgefit}. Uncertainties in $R_e$ and $M_{*, \rm MW}$ are based on the magnification uncertainty; those in the dynamical masses $M_{\rm bulge}$ and $M_{\rm stars}$ also include the $1\sigma$ statistical uncertainty added in quadrature. Local relic galaxies often exhibit a decline in ellipticity at large radii $\gtrsim 5-10$~kpc, as round outer envelope begins to dominate over the disk, and it is debated whether this envelope should be considered part of the bulge \cite{Yildirim15,Savorgnan16}. We found no clear evidence of a declining ellipticity in the outer parts of MRG-M0138 to a limiting distance of $\sim$~12~kpc, measured as the semi-major axis of the isophote, so did not attempt to model any additional component.

We define $\sigma_e^2$ as the intensity-weighted $V_{\rm rms}^2$ within apertures corresponding to three different conventions: i) the half-light ellipse of the entire system, $\sigma_e = 377 \pm 11$~\kms, ii) the half-light ellipse of the bulge, $\sigma_e = 398 \pm 12$~\kms, and iii) a narrow major-axis slit extending to the total half-light radius, $\sigma_e = 430 \pm 13$~\kms. These errors include the 3\% systematic uncertainty derived above. We did not exclude the sphere of influence \cite{McConnell13} and calculate that doing so would have a small effect ($\lesssim 8$~\kms). We used definition ii) to compute $r_{\rm inf} = G M_{\bullet} / \sigma_e^2 = 164$~pc ($G$ is the gravitational constant). Fig.~\ref{fig3} shows MRG-M0138 with the value from definition ii), and the range from definitions i) to iii).

\subsubsection*{Dynamical models: Mass components, parameters, and priors}

We constructed multiple dynamical models using {\tt jampy} version 8.0.0 \cite{Cappellari08,Cappellari20,jampy}. This code calculates stellar dynamics within an axisymmetric system in which the luminosity and mass distributions are specified as MGEs. It assumes that the velocity ellipsoid is aligned with either cylindrical (cyl) or spherical (sph) coordinates; we verify the robustness of the results by the recommended practice \cite{Cappellari20} of considering both these limiting cases.

We modeled three mass components: the stellar system, whose treatment is described below; a dark matter halo; and, optionally, a black hole with a mass $M_{\bullet}$. The dark matter was treated as a spherical halo quantified by a logarithmic slope $\gamma_{\rm DM}$, where the density $\rho_{\rm DM} \propto r^{-\gamma_{\rm DM}}$, and the dark matter fraction $f_{\rm DM}$ within a 5-kpc sphere. The velocity anisotropy parameter $\beta_z$ (cyl models) or $\beta_r$ (sph models) was allowed to vary spatially using the logistic profile in {\tt jampy}. The parameters specifying this gradient are the inner ($\beta_0$) and outer ($\beta_{\infty}$) values of the velocity anisotropy, along with a transition height $z_{\beta}$ (cyl) or radius $r_{\beta}$ (sph). (We fixed {\tt alpha}, the {\tt jampy} parameter that controls the speed of the transition from $\beta_0$ to $\beta_{\infty}$, to 2 and verified that our analysis is insensitive to this choice.) The inclination $i$ was also included as a parameter. We allowed the systemic velocity $v_{\rm sys}$ to vary in initial fits; based on those results, we fixed $v_{\rm sys} = 1$~\kms~(relative to $z = 1.9480$) in the analysis described below.

Measuring $M_{\bullet}$ requires a flexible model of the stellar mass distribution, to ensure that the uncertainties in $M_{\bullet}$ reflect the full plausible range of stellar mass-to-light ratios. Our fiducial stellar mass surface density map (Fig.~\ref{fig:structure}) accounts for variations in stellar age, metallicity, and dust via the SPS models. (Although these parameters can be difficult to disentangle, our mass modeling requires only $M_*/L$, which is the most tightly constrained parameter \cite{Taylor11}.) However, the SPS models assumed a Milky Way IMF, which might not apply to this galaxy, as discussed below. We therefore allowed the stellar mass distribution to deviate from the fiducial map by defining a parameter $\alpha$ such that the dynamical $M_*/L = \alpha \times (M_*/L)_{\rm MW}$, where $(M_*/L)_{\rm MW}$ is the fiducial SPS-based estimate. Although this flexibility is intended primarily to model uncertainty in the IMF, $\alpha$ absorbs any differences between the dynamical and SPS-based stellar mass distributions, regardless of their origin. We introduced further flexibility by allowing $\alpha$ to vary spatially, motivated by evidence for IMF gradients at low redshift (see below). Because we lack prior knowledge of whether these gradients were already in place at $z=2$ and what form they took, we considered several models:

\emph{Model A}: The bulge and disk stars have distinct $\alpha$ values, $\alpha_{\rm bulge}$ and $\alpha_{\rm disk}$, which are spatially constant within each component, i.e., no radial gradients.

\emph{Model B}: The bulge and disk stars participate in a common radial gradient, in which $\alpha$ follows a logistic profile in log $r$ defined by the parameters $\alpha_0$, $\alpha_{\infty}$, $r_{\alpha}^{\rm in}$ and $r_{\alpha}^{\rm out}$:
\begin{equation}
\alpha(r) = \alpha_0 + \frac{\alpha_{\infty} - \alpha_0}{1 + e^{-x}} ~~~~{\rm with}~~~~ x = -2 + 4 \frac{\log r/r_{\alpha}^{\rm in}}{\log r_{\alpha}^{\rm out}/r^{\rm in}_{\alpha}}.
\end{equation}
Thus $\alpha(r) \approx \alpha_0$ at small $r \ll r^{\rm in}_{\alpha}$, $\alpha(r) \approx \alpha_{\infty}$ at large $r \gg r^{\rm out}_{\alpha}$, and $\alpha(r)$ transitions between these values over intermediate radii. We defined $r^{\rm in}_{\alpha}$ and $r^{\rm out}_{\alpha}$ such that the transition is $(1 + e^{-2})^{-1} \approx 88\%$ complete at these radii.

\emph{Model C}: The disk stars participate in the $\alpha$ gradient, separately from a spatially constant $\alpha_{\rm bulge}$.

The adopted priors for all parameters are listed in Table~\ref{tab:posteriors}. Some of the priors are broad and uninformative; here we discuss the informative priors. We limited the inclination to enforce intrinsic ellipticities $\epsilon_{\rm intr} < 0.95$. The velocity anisotropy was limited to avoid non-physical JAM models that are expected when $\beta_z, \beta_r \gtrsim 0.7 \epsilon_{\rm intr} \approx 0.6$ \cite{Wang21}. We limited $\gamma_{\rm DM}$ to the range justified by previous work \cite{Cappellari12}. The most influential priors are those describing $\alpha$. To set these priors, we referred to studies of massive early-type galaxies in the local Universe, a population that we expect MRG-M0138 will evolve to join. Analyses of surface gravity-sensitive stellar absorption lines in such galaxies have shown a bottom-heavy IMF, on average. The most extreme cases reach central values $\alpha \approx 3$ \cite{Conroy17}, declining to $\alpha \approx 1.1$ at radii $R \gtrsim 0.4 R_e$ \cite{vanDokkum17,Mehrgan24}. Dynamical and lensing methods \cite{Posacki15,Mehrgan24} have similarly found $1 \lesssim \alpha \lesssim 2.5$ in low-redshift early-type galaxies with $\sigma_e \gtrsim 250$~\kms. Informed by these bounds, we set a prior range of [0.9, 3.0] on all $\alpha$ parameters and also required $\alpha_{\infty} \leq \alpha_0$, i.e., $\alpha$ is constant or declining as the radius increases. Because the IMF does not appear to vary appreciably at $R \lesssim 0.1 R_e$ in the aforementioned samples, we fixed $r^{\rm in}_{\alpha} = 270~{\rm pc} \approx 0.1 R_e$ in our analysis. This enforces a requirement that $\alpha$ not vary at very small radii, an assumption that also underlies most $M_{\bullet}$ measurements in local galaxies. We allowed $r^{\rm out}_{\alpha}$ to vary from 540~pc (i.e., the transition from $\alpha_0$ to $\alpha_{\infty}$ must occur over at least a factor of two in radius) to 5~kpc (roughly the outer bound of the kinematic data).

\subsubsection*{Dynamical models: Model computation and inference}

Following previous work \cite{Newman18b}, we traced the source kinematics through the lens mapping to the image plane and constrained the dynamical models using a Bayesian framework.

For each spaxel $j$ in the IFU field, we computed the positions of its corners in the source plane. We then subsampled this polygon using a $3 \times 3$ grid of points in the inner regions, or a single central point in the outer regions. At each point $k$, we used {\tt jampy} to compute $V_{k, {\rm rms}}$ and evaluated the surface brightness $I_k$ of the MGE tracer model. The $V_{\rm rms}^2$ value of spaxel $j$ was then computed as the intensity-weighted mean $\sum I_k V_{k, {\rm rms}}^2 / \sum I_k$, where the sums are over $k$, while the surface brightness of spaxel $j$ was computed as the mean $\langle I_k \rangle$. The resulting image-plane maps of $I V_{\rm rms}^2$ and $I$ were then convolved by the NIRSpec PSF, and for each spatial bin, $V_{\rm rms}^2$ was computed as the mean $\langle I V_{\rm rms}^2 \rangle / \langle I \rangle$ over its constituent spaxels. Because this procedure models PSF convolution and pixelization, we disabled the treatment of these effects by {\tt jampy}. This process results in a model vector $\mathbf{m}$ of $V_{\rm rms}$ values that can be directly compared to the observed vector $\mathbf{o}$.

In the {\tt jampy} calculations, the black hole was modeled as compact Gaussian whose radius {\tt rbh} was set to 1~mas. In our models B and C, we imposed an $\alpha$ gradient on the fiducial stellar mass MGE as follows: we first evaluated the stellar mass density profile $\rho_*(r)$ of the disk MGE components in the equatorial plane. We then multiplied this profile by $\alpha(r)$ and fitted another one-dimensional MGE to the modified density profile. We converted this to a two-dimensional MGE, assigning every component the axial ratio $q_{\rm disk}$. In model B, we repeated the same procedure for the bulge component.

We sampled the posterior probability distributions and computed the Bayesian evidence $Z$ using {\tt UltraNest} version 4.3.3. We used a Gaussian likelihood $L \propto \exp(-\frac{1}{2} \chi^2) = \exp[-\frac{1}{2} (\mathbf{o} - \mathbf{m})^{\rm T} {\mathbf \Sigma}^{-1} (\mathbf{o} - \mathbf{m})]$, where $\mathbf \Sigma$ is a covariance matrix (see below) that accounts for the correlations induced by resampling of the IFU data cubes. We masked 9 of the 219 spatial bins due to dust (see above), local mismatch with the MGE model (see above), or proximity to the supernova.

\subsubsection*{Dynamical modeling: Black hole mass and uncertainties}

\begin{figure} 
	\centering
	\includegraphics[width=3in]{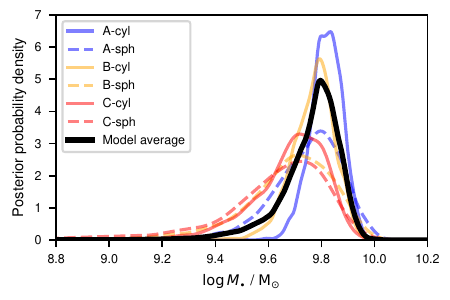}
	\caption{\textbf{Marginalized posterior probability distributions for $M_{\bullet}$.} Colored lines are the marginalized posterior for the black hole mass in each dynamical model (see legend). The thick black line is the same quantity for the Bayesian model average, from which our final estimate is derived.}
    \label{fig:mbh_posterior}
\end{figure}

For each parameterization of the stellar mass distribution (A, B, C), we built both cyl and sph models, resulting in a total of six models (denoted A-cyl, A-sph, etc.). The posteriors are summarized in Table~\ref{tab:posteriors}, and the marginalized posteriors for $M_{\bullet}$ are plotted in Fig.~\ref{fig:mbh_posterior}. We consider the B-cyl model as the fiducial model for purposes of discussion and as a baseline for estimating systematic uncertainties, but our final $M_{\bullet}$ estimate combines results from all six models.

The A-cyl, B-cyl, and C-cyl models produce indistinguishable values of the evidence $Z$ (Table~\ref{tab:posteriors}). They have similar goodness of fit, with a minimum $\chi^2 \approx 364$ for 210 data points and a reduced $\chi^2_{\nu} = 1.7$, which is typical for JAM models \cite{Thater19}. The sph models are disfavored compared to their cyl counterparts by evidence ratios between 11 and 23. All models produce compatible constraints on $M_{\bullet}$ (Fig.~\ref{fig:mbh_posterior}), and we therefore combined their posterior probability distributions using Bayesian model averaging. It is not clear a priori what form any IMF gradients may take in high-redshift galaxies, nor which alignment of the velocity ellipsoid best approximates a multi-component system with a disk and a bulge. We therefore considered all models to have equal prior probability. The model averaging thus amounts to weighting the posteriors by the Bayesian evidence of each model. The averaged posterior indicates $\log(M_{\bullet} / {\rm M}_{\odot}) = 9.78^{+0.08}_{-0.12} {}^{+0.14}_{-0.34}$, where the two uncertainties are the 68\% and 95\% credible intervals.

\begin{figure} 
	\centering
	\includegraphics[width=\textwidth]{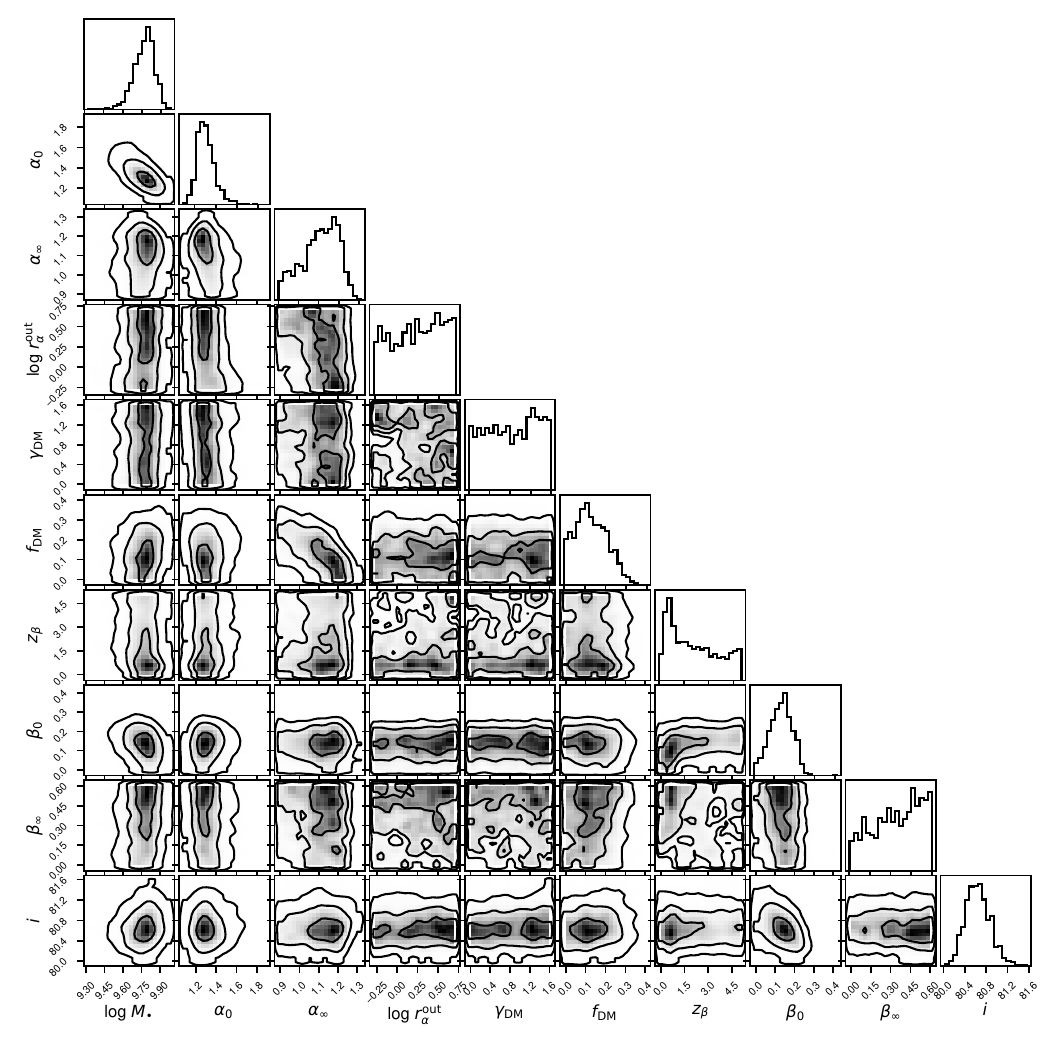}
	\caption{\textbf{Posterior probability distributions for the fiducial dynamical model B-cyl.} For every pair of parameters, a panel with greyscale and contours ($1\sigma$, $2\sigma$, and $3\sigma$) shows the two-dimensional posterior probability density, obtained by marginalizing over all other parameters. Panels at the top of each column show the one-dimensional posterior probability density after marginalizing over all but one parameter. See Table~\ref{tab:posteriors} for units.}
    \label{fig:corner}
\end{figure}

The weak sensitivity of $M_{\bullet}$ to the stellar mass model and the velocity ellipsoid alignment are evidence of its robustness. To further test this robustness, we consider the covariance of $M_{\bullet}$ with other parameters and whether the other parameters  have reasonable values. Considering first the parameters unrelated to $\alpha$, we found that they have little covariance with $M_{\bullet}$ (Fig.~\ref{fig:corner}). All models yielded only an upper limit on $f_{\rm DM}$ (Table~\ref{tab:posteriors}) and no constraint on $\gamma_{\rm DM}$. All models indicated a velocity ellipsoid that is only mildly anisotropic in the center ($\beta_0 \approx 0.1-0.2$) and in some cases increases with radius, although the outer anisotropy is not tightly constrained. Covariance between $M_{\bullet}$ and anisotropy or inclination is minimal. The main covariance is between $M_{\bullet}$ and the value of $\alpha$ near the center. Model A-cyl indicates little spatial variation in $\alpha$, with $\alpha_{\rm disk} \approx \alpha_{\rm bulge} \approx 1.2$. Model B-cyl similarly indicates only a mild $\alpha$ gradient, if any ($\alpha_0 = 1.30^{+0.09}_{-0.07}$ and $\alpha_{\infty} = 1.12^{+0.08}_{-0.12}$). In model C-cyl, a greater range of gradients is permitted in the disk ($\alpha_0 = 1.79^{+0.57}_{-0.44}$ and $\alpha_{\infty} = 1.16^{+0.20}_{-0.17}$). This occurs because the inner disk is less well resolved than the inner bulge: it is much flatter, and its major axis is close to the direction of minimum magnification. Therefore decoupling the inner disk and bulge permits a wider range of $\alpha$ in the former. Model C therefore provides the most stringent test of the robustness of $M_{\bullet}$ to the inner stellar density.

Several systematic uncertainties were evaluated by changing inputs to the fiducial model B-cyl and computing the resulting change to the median of the $M_{\bullet}$ posterior: i) To estimate the effect of uncertainties in the magnification, we isotropically rescaled the source plane coordinates and MGE dimensions by linear factors of 1.28 and 0.83, which correspond to $-39\%$ and +44\% changes to $\mu$. This shifted $M_{\bullet}$ by +0.10~dex and $-0.08$~dex, respectively. ii) We used the alternative PSF A and found a negligible 0.01~dex shift. iii) We uniformly changed all $\sigma$ measurements by our estimated systematic uncertainty of $\pm3$\%, which shifted $M_{\bullet}$ by ${}^{+0.03}_{-0.04}$~dex. Adding these components in quadrature, we estimate a systematic uncertainty of ${}^{+0.11}_{-0.09}$~dex in $M_{\bullet}$.

Our dynamical models neglect the mass of gas and dust. Dust continuum emission at 1.3~mm has been observed in MRG-M0138 \cite{Whitaker21}, resulting in a low estimated gas-to-stellar mass fraction of 0.6\% and a gas mass, given our adopted magnification, of $\approx1.3 \times 10^9$\msol. The millimeter observations do not resolve the gas distribution, so we estimate it from our maps of the dust attenuation $A_V$. If the neutral gas surface density is roughly proportional to $A_V$, then $\sim$2\% is projected within $r_{\rm inf}$. Therefore, although we do not include gas in the systematic error budget, we expect it to be dynamically negligible. In the extreme limiting case in which the gas mass is fully contained within $r_{\rm inf}$ and is subtracted from our $M_{\bullet}$ estimate, the result would change by $<1\sigma$.

\subsubsection*{Dynamical modeling: Necessity of a black hole}

\begin{figure} 
	\centering
	\includegraphics[width=4.25in]{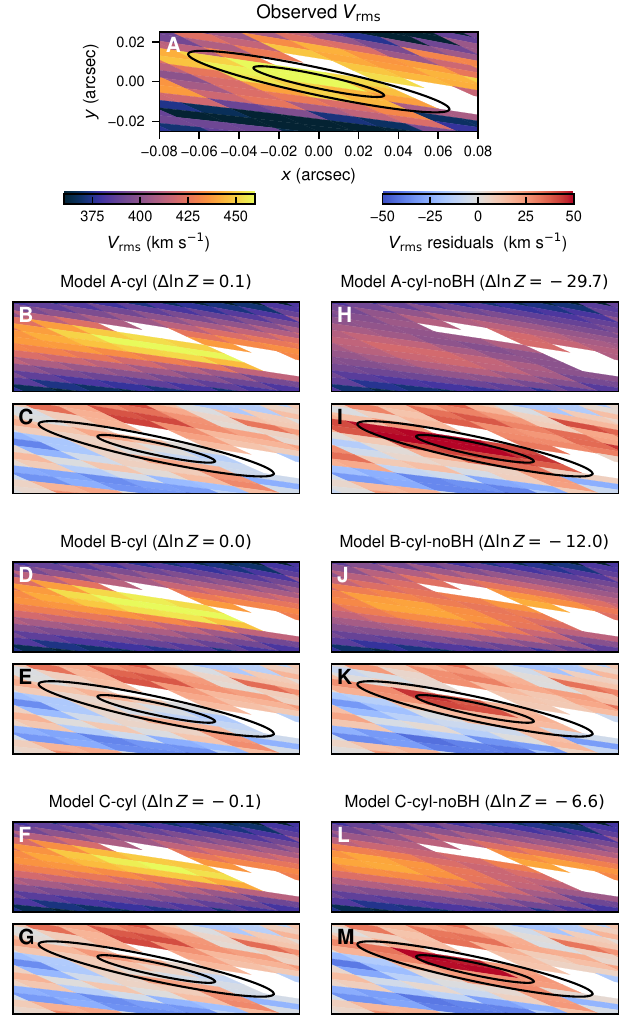}
	\caption{\textbf{Comparison of the observed stellar kinematics near the galaxy center to dynamical models with and without a black hole.} \textbf{(A)} The observed $V_{\rm rms}$ in the source plane. Ellipses centered on the galaxy nucleus show the PSF (dimensions $1 \sigma_{\rm psf}$ and $2 \sigma_{\rm psf}$). \textbf{(B)} The  maximum likelihood $V_{\rm rms}$ field for the dynamical model A-cyl. \textbf{(C)} The $V_{\rm rms}$ residuals for model A-cyl, defined as the difference between panels A and B. Ellipses are the same as in panel A. \textbf{(D-E}) Same as panels B-C but for model B-cyl. \textbf{(F-G}) Same as panels B-C but for model C-cyl. \textbf{(H-M}) Same as panels B-G but for the corresponding models without a black hole. The models lacking a black hole do not match the central $V_{\rm rms}$ peak. Each panel is labeled with the corresponding value of $\Delta \ln Z$, the log Bayesian evidence ratio compared to the fiducial model (B-cyl). White regions were masked due to dust or the supernova.
    \label{fig:compare_bh_nobh}}
\end{figure}

Having determined that a range of models consistently indicate the presence of a black hole (Fig.~\ref{fig:mbh_posterior}), we next evaluated the extent to which a black hole is demanded by the observations and considered alternative explanations. We augmented models A-cyl, B-cyl, and C-cyl by building counterparts that do not include a black hole, which are designated A-cyl-noBH, etc. Fig.~\ref{fig:compare_bh_nobh} compares the observed $V_{\rm rms}$ field in the center of MRG-M0138 to these six models. The high central $V_{\rm rms}$ is not reproduced by any of the models that lack a black hole (Fig.~\ref{fig:compare_bh_nobh}H-M). This is also demonstrated in Fig.~2 for the models B-cyl, B-cyl-noBH, and A-cyl-noBH. Model A-cyl-noBH, which lacks radial IMF gradients, fares the worst. Models B-cyl-noBH and C-cyl-noBH attempt to compensate for the lack of a central dark mass with IMF gradients, but they still do not match the data. Quantitatively, we compared models using the log evidence ratio, $\Delta \ln Z$, relative to the fiducial model (B-cyl). The no-black-hole models have much lower evidence, $\Delta \ln Z = -29.7$ to $-6.6$ (Fig.~\ref{fig:compare_bh_nobh}), i.e., they are disfavored by Bayes factors of 735 to $8 \times 10^{12}$.

Although our models were designed to be realistic, simpler models can also fit the data. We tested a very simple dynamical model that lacks dark matter and gradients in $\alpha$ or $\beta_z$ so has only four parameters, $\alpha$, $\beta_z$, $M_{\bullet}$, and $i$. It is a limiting case of the more complex models A-cyl, B-cyl, and C-cyl, and we found that its evidence is moderately higher, by $\Delta \ln Z \approx 2$. This indicates that the increased complexity of models A-C is not demanded by the data, whereas a black hole remains a necessary component. We chose not to include this simple model in our averaging because of the strong assumptions it requires; it indicated a consistent $\log(M_{\bullet} / {\rm M}_{\odot}) = 9.81^{+0.05}_{-0.06}$.

\begin{figure} 
	\centering
	\includegraphics[width=3in]{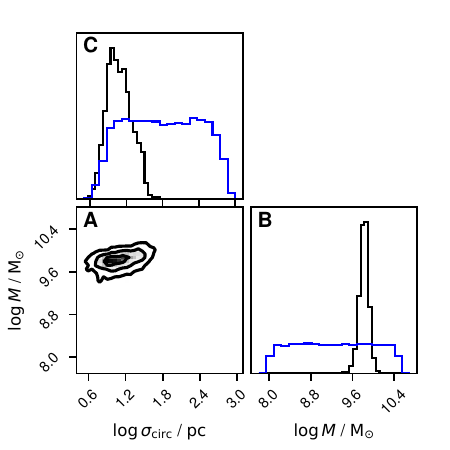}
	\caption{\textbf{Constraints on the mass and size of a central dark mass.}    
    Posterior probability densities are shown for a modified version of the fiducial dynamical model, in which the black hole is replaced by a central dark mass with a circularized size $\sigma_{\rm circ}$ and mass $M$. \textbf{(A)} The joint probability density for $M$ and $\sigma_{\rm circ}$ is shown in greyscale with $1\sigma$, $2\sigma$, and $3\sigma$ contours. \textbf{(B)} The black histogram is the marginalized probability density for $M$. The blue histogram is the prior. \textbf{(C)} Like B but for $\sigma_{\rm circ}$. In this case, the blue histogram is an effective prior arising from the priors on the semi-major axis $\sigma$ and axial ratio $q$. The $\sigma_{\rm circ}$ posterior is prior-limited at the low end, indicating an unresolved mass.\label{fig:dark_mass}}
\end{figure}

We further considered whether the observed kinematics can plausibly be explained by a central dark mass other than a black hole. We elaborated the fiducial model B-cyl by replacing the black hole with an extended dark mass, a single Gaussian component defined by its mass $M$, size $\sigma$ (major axis), and projected axial ratio $q$. We placed a log-uniform prior on $\sigma$ ranging from $0.001''$ to $0.1'' \approx 86$~pc, a log-uniform prior on $M / {\rm M}_{\odot}$ from $10^8 - 10^{10.5}$, and a uniform prior on $q$ from $q_{\rm disk}$ to 1. We found that an extremely compact and dense mass is required to fit the kinematics, as shown in Fig.~\ref{fig:dark_mass}. The circularized size $\sigma_{\rm circ} = \sigma q^{1/2} < 28$~pc (95\% upper limit), while the required mass is $10^{9.83 \pm 0.09}$\msol~and the average surface density within $\sigma$ is $10^{6.8 \pm 0.4}$\msol~pc$^{-2}$. If this dark mass corresponds to a stellar system, it would have an extreme surface density compared to observed stellar systems \cite{Baggen24} and exceed theoretical limits by an order of magnitude \cite{Grudic18}. To remain hidden in NIRCam imaging out to $\lambda_{\rm rest} = 1.5$~µm, the stars would have to be highly obscured, despite the low attenuation $A_V \approx 0.2$~mag that we measure toward the galaxy center. We consider this scenario very implausible compared to a black hole.

Our dynamical models are all axisymmetric and cannot directly address the possible influence of a bar on our measurements. We do not see clear evidence of a bar in the images. It has been suggested that a thin nuclear bar, if it is viewed end-on, might remain hidden and boost the line-of-sight stellar motion \cite{Gerhard88,Emsellem13}. Although we cannot exclude this hidden bar scenario, we consider it unlikely: it requires a specific viewing angle; MRG-M0138 is a very dense galaxy with a fairly high bulge fraction, characteristics that are expected to inhibit bar formation \cite{Kataria18}; and the position of the MRG-M0138 black hole on the scaling relations is consistent with relic galaxies (as discussed in the main text).

\subsubsection*{AGN luminosity and Eddington ratio limits}

We assessed the H$\alpha$ luminosity $L_{\rm H\alpha}$ in the central spaxel, corrected for aperture losses assuming point-like emission, to minimize the contribution from non-nuclear sources. To improve sensitivity to very weak line emission, we used $N_{\rm mult} = 31$ in the SPS fitting, corresponding to one polynomial order per $10^4$~\kms. We found correlated residuals to the fitted SPS models of $\pm1.5$\%, corresponding to the amplitude of a Gaussian emission line with a rest-frame equivalent width ${\rm EW}_0 = 0.3$~\AA~at H$\alpha$. To account for this systematic, we added 0.3~\AA~in quadrature to the statistical uncertainties in ${\rm EW}_0$ (derived from the {\tt ppxf} flux uncertainties) and likewise increased the $L_{\rm H\alpha}$ uncertainties. 

\begin{figure}
    \centering
    \includegraphics[width=4in]{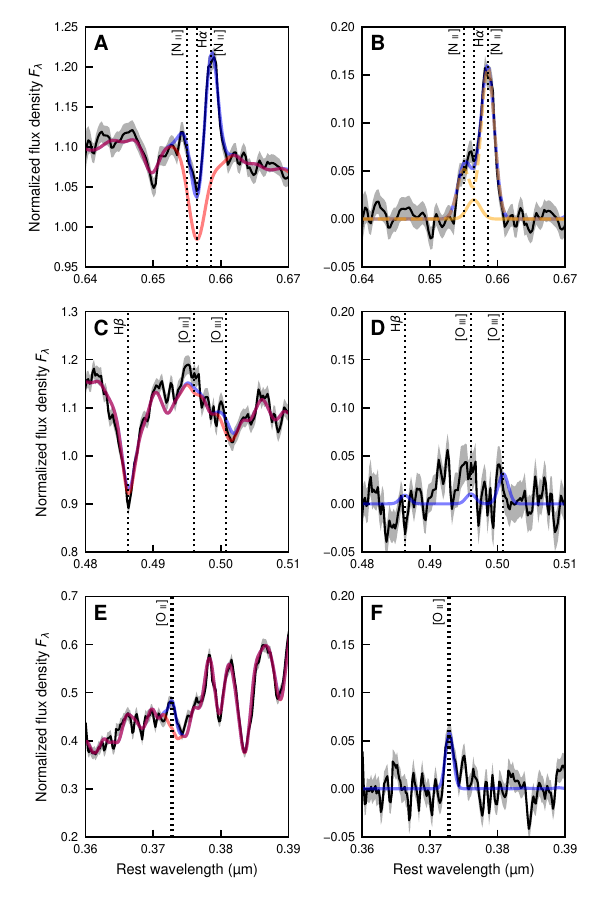}
    \caption{\textbf{Nuclear spectrum constraining gaseous line emission.} {\bf (A)} The black curve (with $\pm 1 \sigma$ band) shows the observed spectrum in the vicinity of H$\alpha$ and [N~\textsc{ii}]. The red curve shows the stellar continuum of the best-fitting model. The blue curve shows the best-fitting model including the line emission. Dotted lines show the rest-frame wavelengths of the labeled transitions. {\bf (B)} The black curve shows the observed spectrum with the stellar continuum of the best-fitting model subtracted. The blue curve shows the total emission line model, while the orange curves separate the H$\alpha$ (solid) and [N~\textsc{ii}] (dashed) components. {\bf (C-D)} Same as panels A-B, but showing the wavelength range around H$\beta$ and [O~\textsc{iii}]. The lines are not blended, so we do not show their individual contributions. {\bf (E-F)} Same as panels C-D, but showing the wavelength range around [O~\textsc{ii}]. Only the [N~\textsc{ii}] and [O~\textsc{ii}] lines are detected, at significances $11\sigma$ and $5\sigma$, respectively; we derive limits on the others.
    \label{fig:emlinespec}}
\end{figure}

We detected the [N~\textsc{ii}]$\lambda\lambda$6550,6585 and [O~\textsc{ii}]$\lambda\lambda3727,3730$ emission lines (Fig.~\ref{fig:emlinespec}) at significances of $5\sigma$ and $11\sigma$, respectively. No other emission line was detected with $> 2\sigma$ significance. We placed $2\sigma$ upper limits on the H$\alpha$ ${\rm EW}_0 < 1.1$~\AA~and $L_{\rm H\alpha} < 10^{39.6}$~erg~s${}^{-1}$, which we converted to a limit on $L_{\rm bol} < 10^{42.1}$~erg~s${}^{-1}$ assuming a bolometric correction \cite{Ho09} $C_{\rm H\alpha} = L_{\rm bol} / L_{\rm H\alpha} = 300$. These estimates reflect our detection limits, but there are additional systematic uncertainties. Because we cannot constrain the Balmer decrement, we assumed that the emission lines are attenuated like the starlight; at H$\alpha$, this attenuation is $A_{\rm H\alpha} = 0.2$~mag. Emission lines in local LINER nuclei have low extinction; the distribution of color excess $E(B-V)$ has a $2\sigma$ upper limit of 0.5~mag \cite{Ho03}. Based on this, we increased our upper limits to allow for possible additional extinction of $A_{\rm H\alpha} = 1.0$~mag and for 0.3~dex uncertainly in $C_{\rm H\alpha}$ \cite{Ho09}, arriving at $L_{\rm bol} < 10^{42.8}$~erg~s${}^{-1}$ and $\lambda_{\rm Edd} < 10^{-5.1}$.

We constrained the X-ray luminosity using an archival 25~ks observation (Obs ID 17186) by the Advanced CCD Imaging Spectrometer (ACIS) on the Chandra X-ray Observatory. We used {\tt CIAO} version 4.17 \cite{CIAO,ciaoascl} to estimate a credible interval on the model flux for point-like emission at the center of MRG-M0138, in the multiple image that we observed with NIRSpec. The spectral model was a power law with a photon index $\Gamma = 1.8$ multiplied by foreground Milky Way absorption, initially assuming negligible internal absorption. We fitted the model over the observed-frame energy range 0.5 to 7~keV then corrected the model fluxes to refer to the rest-frame 2 to 10~keV range. The 90\% credible interval on the demagnified X-ray luminosity is $L_X = 0-10^{42.2}$~erg~s${}^{-1}$. We assumed a bolometric correction factor $C_X = 15.8$ \cite{Ho09}, similar to the $C_X \approx 9$ previously found for AGNs with $L_X$ equal to our upper limit \cite{Netzer19}. The 90\% credible interval for $L_{\rm bol} = C_X L_X = 0-10^{43.4}$~erg~s${}^{-1}$. The X-ray non-detection implies $\lambda_{\rm Edd} < 10^{-4.5}$, which is consistent with but less constraining than H$\alpha$. We investigated the possible consequences of a Compton-thick absorber by adding internal absorption to the spectral model. For an assumed column density of $\log(N_H / {\rm cm}^{-2}) = 24$ or 25, we find limits of $\lambda_{\rm Edd} < 10^{-3.6}$ or $10^{-2.2}$, respectively. Because the redshift allows Chandra to access hard X-rays, we conclude that the non-detection rules out an active black hole ($\lambda_{\rm Edd} \gtrsim 0.01$), even if it is quite obscured.

\subsubsection*{Construction of Figures 2 and 3}

Fig.~\ref{fig2} shows kinematics along slits aligned with the major, intermediate, and minor axes. The kinematics can undergo strong gradients among and within spatial bins, leading to an artificially jagged appearance of the one-dimensional slit kinematics that is due to the irregular Voronoi binning scheme. We addressed this as follows. Model curves in each panel were produced from highly sampled kinematic fields that were PSF-convolved, but not Voronoi binned. The data points included in each panel are those that spatially overlap with a narrow (8~mas wide) slit placed along the specified axis. We used the fiducial dynamical model to derive a correction for each bin, equal to the difference between the smooth model curve (described above) and the Voronoi-binned model $V_{\rm rms}$. We added this correction to the observed $V_{\rm rms}$ for display purposes. This method preserves the residuals between the data and the fiducial model, allowing us to judge the goodness of fit while also showing the underlying smooth kinematics. Regions contaminated by the supernova were excluded from the kinematic measurements and are omitted Fig.~\ref{fig2}B; additional bins masked for dynamical modeling are omitted in Fig.~\ref{fig2}A.

In Fig.~\ref{fig3}A, for some of the $z \sim 1$ to 3 quasars \cite{Shen16}, we followed previous work to compute $L_{\rm bol}$ \cite{Shen11} and $M_{\bullet}$ \cite{Bongiorno14}. For the $z \sim 2$ and 5 quiescent galaxies, we adopted uncertainties of 0.6~dex \cite{Vestergaard06} in the broad-line estimates of $M_{\bullet}$; for the $z \sim 5$ quiescent galaxy \cite{Carnall23}, we estimated $L_{\rm bol}$ following previous work \cite{Greene05,Shen11}.

\subsubsection*{Kinematic covariance matrix}
 
The {\tt emsm} resampling method computes each voxel in the output data cube as a weighted average of pixels in our four dithered exposures. The same input pixel receives appreciable weight in neighboring output voxels, particularly when resampling onto a smaller-than-native pixel scale as we do, and therefore noise is correlated in nearby spaxels. Here we derive the covariance matrix $\mathbf \Sigma$ that describes our $V_{\rm rms}$ measurements.

 The {\tt emsm} weights for our configuration are $w = \exp[-(R / 0.05'')^2]$, where $R = [(\Delta x)^2 + (\Delta y)^2]^{0.5}$, and $\Delta x$ and $\Delta y$ are the separation between a given input pixel and a given output voxel. $w = 0$ beyond a region of interest specified by the radius $R_{\rm ROI} = 0.2''$. (The weights and $R_{\rm ROI}$ used in {\tt emsm} resampling vary with wavelength; we used values near the mid-point of our observed wavelength range.)
 
 We first used numerical simulation to estimate the flux correlations among nearby output spaxels within the same wavelength slice. Beginning with exposure one, we considered a set of input pixels that map onto the wavelength slice of interest. We consider these pixels to map onto a regular grid in the output plane with a spacing of $0.1''$, which is the native pixel scale and slice width. This assumption neglects distortion, which is locally negligible, and for simplicity we also neglected the rotation of the output grid with respect to instrument coordinates. We populated these pixels with Gaussian random noise. We likewise generated random noise for pixels in exposures two to four; these map onto output spaxel grids that are shifted by $\pm 0.05''$ in each axis, reflecting the dither pattern. We then constructed the output image with a spaxel size of $0.05''$, using the {\tt emsm} weights described above to average the input pixels. We computed the correlation coefficient $\rho_{\Delta x', \Delta y'}$ as a function of the absolute separation $(\Delta x', \Delta y')$ of the output spaxel coordinates. The result is only weakly sensitive to the phase of the output grid with respect to the input grids; we therefore marginalized over phase to derive $\rho_{0,0} = 1$, $\rho_{0,1} = \rho_{1,0} = 0.61$, $\rho_{1,1} = 0.37$, $\rho_{0,2} = \rho_{2,0} = 0.14$, $\rho_{2,1} = \rho_{1,2} = 0.08$, and $\rho_{2,2} = 0.02$. For $\Delta x', \Delta y' > 2$, we approximate $\rho_{\Delta x', \Delta y'} = 0$.

This matrix describes the spatial correlations among the fluxes in each wavelength slice of the output data cube. It also describes correlations among $V_{\rm rms}$ values extracted from spectra in nearby output spaxels. With correlations thus determined at the spaxel level, we estimated the covariance $\Sigma_{A,B}$ between $V_{\rm rms}$ measures extracted from two spatially binned spectra, enumerated $A$ and $B$:
\begin{equation}
    \Sigma_{A, B} = \frac{\sum\limits_{(j,k) \in A, (l,m) \in B} \rho_{|j-k|, |l-m|}}{\left(\sum\limits_{(j,k) \in A, (l,m) \in A} \rho_{|j-k|, |l-m|}\right) \left(\sum\limits_{(j,k) \in B, (l,m) \in B} \rho_{|j-k|, |l-m|}\right)} \sigma^{\rm rms}_A \sigma^{\rm rms}_B,
    \label{eqn:bincorrs}
\end{equation}
where $\sigma^{\rm rms}_A$ and $\sigma^{\rm rms}_B$ are our estimated uncertainties in $V_{\rm rms}$ in bins A and B, respectively, and $(j,k)$ and $(l,m)$ enumerate the coordinates of the spaxels comprising a spatial bin. To derive this expression, we assumed linearity and  neglected flux and $\sigma^{\rm rms}$ differences among spaxels within a bin. We expect these gradients to be small within small spatial bins. Those bins are the ones of main interest, because large bins composed of many spaxels are essentially uncorrelated with other bins anyway. We used the $\mathbf \Sigma$ matrix to compute likelihoods in our dynamical modeling.



\clearpage

\clearpage

\begin{table} 
	\centering
	\caption{\small \textbf{Multi-Gaussian expansions of the stellar light and mass distributions.} Each pair of peak surface brightness and $\sigma_{\rm MGE}$ in the left two columns specify one Gaussian component. $\sigma_{\rm MGE}$ is the size measured along the major axis, which has a position angle of $215.7^{\circ}$. The components are grouped into the bulge and disk based on their axial ratio. The right two columns similarly specify components of the multi-Gaussian expansion of the stellar mass surface density.}
	\label{tab:mge} 

	\begin{tabular}{cc|cc}
		\\
		\hline
	      Peak surface & $\sigma_{\rm MGE}$ (arcsec) &
        Peak surface & $\sigma_{\rm MGE}$ (arcsec) \\
        brightness (${\rm L}_{\odot}$ pc${}^{-2}$) & & density (${\rm M}_{\odot}$ pc${}^{-2}$) & \\
		\hline
		\multicolumn{4}{c}{Bulge ($q_{\rm bulge} = 0.600$)} \\
		\hline
        $4.20 \times 10^4$ & 0.0404 & $5.05 \times 10^4$ & 0.0291 \\
        $4.06 \times 10^4$ & 0.0585 & $5.28 \times 10^4$ & 0.0571 \\
        $4.14 \times 10^2$ & 0.286 & $1.76 \times 10^2$ & 0.734 \\
        $2.12 \times 10^2$ & 0.690 & & \\
        \hline
		\multicolumn{4}{c}{Disk ($q_{\rm disk} = 0.183$)} \\
		\hline
        $3.93 \times 10^5$ & 0.00713 & $4.43 \times 10^5$ & 0.00700 \\
        $9.57 \times 10^4$ & 0.0323 & $5.71 \times 10^4$ & 0.0240 \\
        $1.61 \times 10^3$ & 0.111 & $3.52 \times 10^4$ & 0.0373 \\
        $1.23 \times 10^4$ & 0.255 & $2.72 \times 10^3$ & 0.127 \\
        $1.82 \times 10^3$ & 0.500 & $7.66 \times 10^3$ & 0.273 \\
        $1.04 \times 10^3$ & 0.615 &$2.18 \times 10^3$ & 0.549 \\
        \hline
	\end{tabular}
    
\end{table}

\begin{table} 
	\centering
	\caption{\small \textbf{Parameter constraints from dynamical modeling}. For each parameter in the six dynamical models, we list the assumed prior and the marginalized posterior: for $f_{\rm DM}$ and $\beta_{\infty}$, 95\% limits are given; for $r^{\rm out}_{\alpha}$ and $\gamma_{\rm DM}$, the posteriors are nearly the same as the priors and so are omitted; and for all other parameters, the median value is listed with uncertainties corresponding to the 16th and 84th percentiles. $\Delta \ln Z$ is the log Bayesian evidence relative to the fiducial model (B-cyl). All priors are uniform between the listed bounds. The lower bounds for $\beta_0$ and $\beta_{\infty}$ are 0 and $-1$ for the cyl and sph models, respectively.}
	\label{tab:posteriors} 

    {\small
	\begin{tabular}{lcccc}
\hline
Parameter & Prior & Units & Posterior (cyl) & Posterior (sph) \\
\hline

\multicolumn{5}{c}{Model A}\\
\hline
$\alpha_{\rm bulge}$ & [0.9, 3] & $\ldots$ & $1.237^{+0.069}_{-0.073}$ & $1.27^{+0.11}_{-0.13}$\\
$\alpha_{\rm disk}$ & [0.9, 3] & $\ldots$ & $1.17^{+0.19}_{-0.17}$ & $1.23^{+0.21}_{-0.19}$\\
log $M_{\bullet}$ & [8, 10.5] & ${\rm M}_{\odot}$ & $9.821^{+0.057}_{-0.060}$ & $9.77^{+0.10}_{-0.15}$\\
$\gamma_{\rm DM}$ & [0, 1.6] & $\ldots$ & $\ldots$ & $\ldots$\\
$f_{\rm DM}$ & [0, 1] & $\ldots$ & $< 0.23$ & $< 0.22$\\
$z_{\beta}$ or $r_{\beta}$ & [0, 5] & kpc & $1.8^{+2.1}_{-1.3}$ & $2.3^{+1.8}_{-1.5}$\\
$\beta_0$ & [0 or -1, 0.6] & $\ldots$ & $0.138^{+0.055}_{-0.067}$ & $0.17^{+0.15}_{-0.18}$\\
$\beta_{\infty}$ & [0 or -1, 0.6] & $\ldots$ & $> 0.060$ & $> -0.72$\\
$i$ & [79.7, 90] & deg & $80.66^{+0.25}_{-0.23}$ & $81.15^{+0.27}_{-0.25}$\\
$\Delta \ln Z$ & $\ldots$ & $\ldots$ & $0.15 \pm 0.69$ & $-2.21 \pm 0.75$ \\
min. $\chi^2$ & $\ldots$ & $\ldots$ & 363.4 & 369.0\\
\hline
\multicolumn{5}{c}{Model B}\\
\hline
$\alpha_0$ & [0.9, 3] & $\ldots$ & $1.299^{+0.085}_{-0.068}$ & $1.359^{+0.117}_{-0.087}$\\
$\alpha_{\infty}$ & [0.9, 3] & $\ldots$ & $1.118^{+0.082}_{-0.115}$ & $1.159^{+0.092}_{-0.121}$\\
log $r_{\alpha}^{\rm out}$ & [log 0.54, log 5] & kpc & $\ldots$ & $\ldots$\\
log $M_{\bullet}$ & [8, 10.5] & ${\rm M}_{\odot}$ & $9.778^{+0.064}_{-0.088}$ & $9.70^{+0.14}_{-0.18}$\\
$\gamma_{\rm DM}$ & [0, 1.6] & $\ldots$ & $\ldots$ & $\ldots$\\
$f_{\rm DM}$ & [0, 1] & $\ldots$ & $< 0.26$ & $< 0.27$\\
$z_{\beta}$ or $r_{\beta}$ & [0, 5] & kpc & $1.8^{+2.1}_{-1.3}$ & $2.3^{+1.8}_{-1.5}$\\
$\beta_0$ & [0 or -1, 0.6] & $\ldots$ & $0.135^{+0.054}_{-0.060}$ & $0.18^{+0.14}_{-0.16}$\\
$\beta_{\infty}$ & [0 or -1, 0.6] & $\ldots$ & $> 0.046$ & $> -0.79$\\
$i$ & [79.7, 90] & deg & $80.64^{+0.24}_{-0.21}$ & $81.15^{+0.28}_{-0.26}$\\
$\Delta \ln Z$ & $\ldots$ & $\ldots$ & 0 & $-3.11 \pm 0.67$ \\
min. $\chi^2$ & $\ldots$ & $\ldots$ & 363.7 & 369.7\\
\hline
\multicolumn{5}{c}{Model C}\\
\hline
$\alpha_0$ & [0.9, 3] & $\ldots$ & $1.79^{+0.57}_{-0.44}$ & $1.90^{+0.60}_{-0.46}$\\
$\alpha_{\infty}$ & [0.9, 3] & $\ldots$ & $1.16^{+0.20}_{-0.17}$ & $1.22^{+0.22}_{-0.20}$\\
log $r_{\alpha}^{\rm out}$ & [log 0.54, log 5] & kpc & $\ldots$ & $\ldots$\\
$\alpha_{\rm bulge}$ & [0.9, 3] & $\ldots$ & $1.137^{+0.094}_{-0.106}$ & $1.14^{+0.15}_{-0.15}$\\
log $M_{\bullet}$ & [8, 10.5] & ${\rm M}_{\odot}$ & $9.70^{+0.11}_{-0.16}$ & $9.66^{+0.14}_{-0.23}$\\
$\gamma_{\rm DM}$ & [0, 1.6] & $\ldots$ & $\ldots$ & $\ldots$\\
$f_{\rm DM}$ & [0, 1] & $\ldots$ & $< 0.24$ & $< 0.23$\\
$z_{\beta}$ or $r_{\beta}$ & [0, 5] & kpc & $1.7^{+2.2}_{-1.2}$ & $2.1^{+1.9}_{-1.3}$\\
$\beta_0$ & [0 or -1, 0.6] & $\ldots$ & $0.121^{+0.057}_{-0.063}$ & $0.12^{+0.16}_{-0.19}$\\
$\beta_{\infty}$ & [0 or -1, 0.6] & $\ldots$ & $> 0.052$ & $> -0.74$\\
$i$ & [79.7, 90] & deg & $80.66^{+0.26}_{-0.21}$ & $81.12^{+0.27}_{-0.25}$\\
$\Delta \ln Z$ & $\ldots$ & $\ldots$ & $-0.13 \pm 0.62$ & $-2.72 \pm 0.70$ \\
min. $\chi^2$ & $\ldots$ & $\ldots$ & 363.5 & 369.3\\
\hline
\end{tabular}}
\end{table}

\begin{table} 
	\centering
	\caption{\small \textbf{Data sources for Fig.~3.}}
	\label{tab:fig3sources} 

	\begin{tabular}{lll}
\hline
Subpanel & Source class in legend & References \\
\hline
A & $z \sim 1$--3 quasars & \cite{Shen16,Bischetti17} \\
A, B & $z \sim 1$--3 BLAGN & \cite{Ding20,Suh20} \\
A & $z \sim 6$ quasars & \cite{Shen19} \\
A & $z > 4$ BLAGN & \cite{Harikane23,Greene24,Juodzbalis24,Maiolino24,Matthee24} \\
A, B, D & $z \sim 2$ \& 5 QGs & \cite{Carnall23,Ito25} \\
A & Local galaxies & \cite{Ho09} \\
B & Local early- and late-type galaxies & \cite{Greene20} \\
B & $z \sim $1--3 quasars & \cite{Decarli10a,Decarli10b}\\
C, D & Local early-type galaxies & \cite{KH13} \\
C, D & Relic galaxies & \cite{Walsh-Mrk1216,Walsh-N1271,Walsh-N1277,Cohn21,Cohn23,Cohn24}\\
\hline
\end{tabular}
\end{table}





\end{document}